%% file: main.tex
\documentclass[sigconf]{acmart}
\usepackage[inline]{enumitem}
\usepackage{textcomp}
\usepackage{longtable}
\usepackage{caption}
\usepackage{multirow}

\usepackage[capitalize, noabbrev]{cleveref}
\usepackage{colortbl}
\usepackage{pgffor}

\date{September 2024}

\newcommand{\studyN}{N=16}
\newcommand{\system}{CorpusStudio}

\usepackage{tikz}
\usepackage{xcolor}
\usepackage{calc}
\usepackage{tabularx}

\usepackage{acmart-taps}
\newcommand{\p}[1]{%
\begingroup%
  \def\temp{}%
  \foreach \i [count=\n from 1] in {#1} {%
    \ifnum\n=1 $P_{\i}$\else\temp, $P_{\i}$\fi
  }%
\endgroup
}

\definecolor{csnoteColor}{rgb}{0.7,0,0.9}
\definecolor{csnoteColor2}{rgb}{0.7,0.5,0.9}

\newcommand{\changenote}[1]{#1} %
\newcommand{\changenoteAcc}[1]{#1}%

\newcommand{\csnote}[1]{#1} 
\newcommand{\userquote}[1]{\textit{#1}}

\copyrightyear{2025}
\acmYear{2025}
\setcopyright{cc}
\setcctype{by}
\acmConference[CHI '25]{CHI Conference on Human Factors in Computing Systems}{April 26-May 1, 2025}{Yokohama, Japan}
\acmBooktitle{CHI Conference on Human Factors in Computing Systems (CHI '25), April 26-May 1, 2025, Yokohama, Japan}\acmDOI{10.1145/3706598.3713974}
\acmISBN{979-8-4007-1394-1/25/04}

\begin{document}

\title[CorpusStudio: Surfacing Emergent Patterns in a Corpus of Prior Work while Writing]{CorpusStudio: Surfacing Emergent Patterns in a Corpus of Prior Work while Writing}

\author{Hai Dang}
\authornote{Equal contributions.}
\email{hai.dang@uni-bayreuth.de}
\orcid{0000-0003-3617-5657}
\affiliation{%
  \institution{University of Bayreuth}
  \country{Germany}
}

\author{Chelse Swoopes}
\authornotemark[1]
\email{cswoopes@g.harvard.edu}
\orcid{0000-0002-7813-8937}
\affiliation{%
  \institution{Harvard University}
  \country{USA}
}

\author{Daniel Buschek}
\orcid{0000-0002-0013-715X}
\email{daniel.buschek@uni-bayreuth.de}
\affiliation{%
  \institution{University of Bayreuth}
  \country{Germany}
}

\author{Elena L. Glassman}
\email{glassman@seas.harvard.edu}
\orcid{0000-0001-5178-3496}
\affiliation{%
  \institution{Harvard University}
  \country{USA}
}

\begin{abstract}  
\input{Sections/abstract_final}

\end{abstract}



\begin{teaserfigure}
  \centering
  \includegraphics[width=0.9\textwidth]{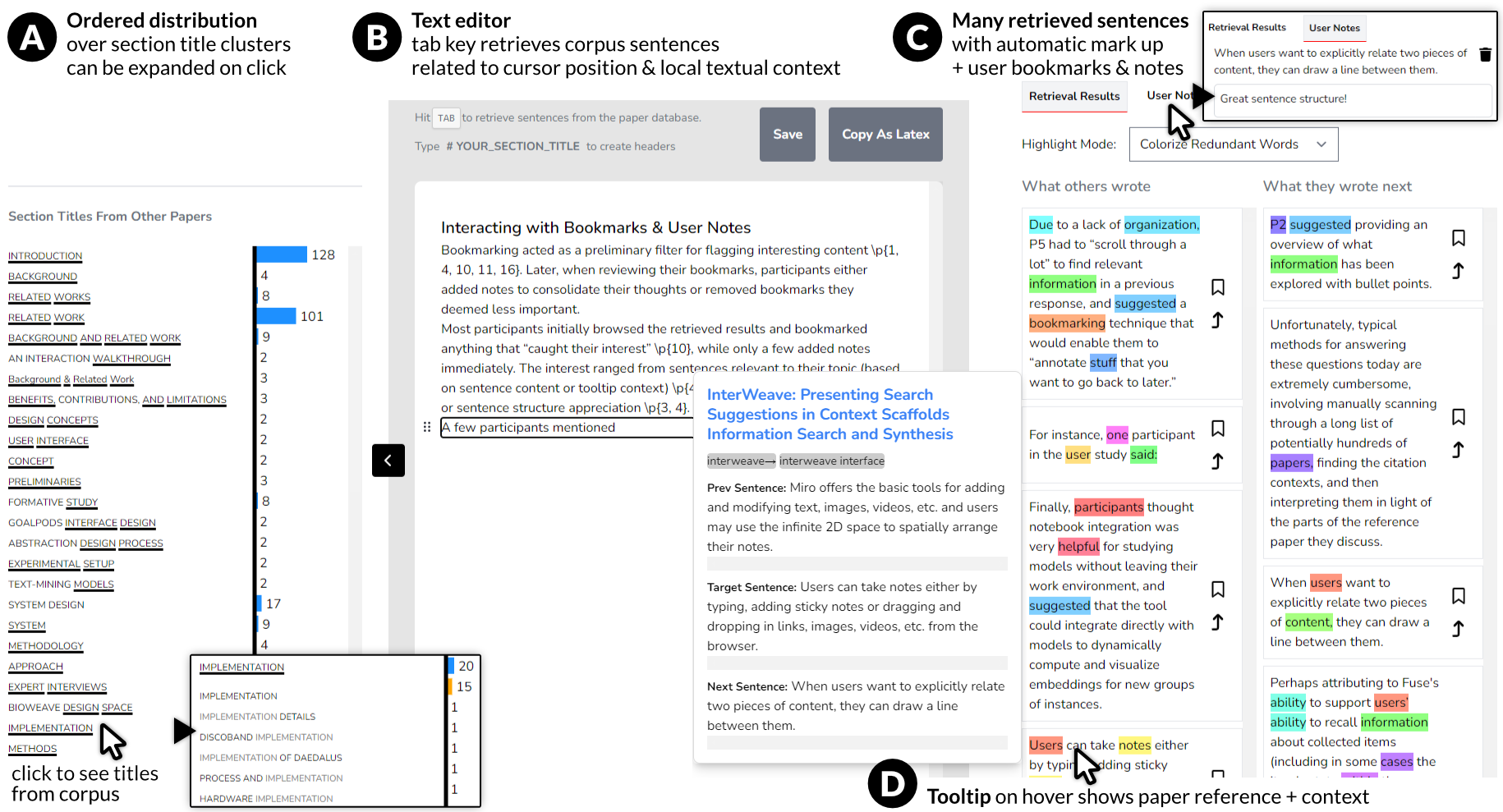}
  \caption{\changenoteAcc{Given a corpus of papers written for the same or similar audiences, e.g., papers previously published at ACM UIST, \system ~supports writers by making visible the writing choices of previous authors in the corpus. To help the writer recognize common and uncommon paper structures, the left sidebar \textit{(A)} shows an ordered distribution of clusters of section titles in the corpus using Positional Diction Clustering~\cite{gero2024supporting}. Informed by \textit{(A)}, writers can draft their own outline in the center text editor \textit{(B)}. When fleshing out their outline with prose, to potentially see emergent patterns in previously written papers, writers can press \texttt{TAB} to retrieve analogous sentence examples from the corpus \textit{(C)} based on their cursor's location within their own draft. To reveal emerging patterns, the writer can select different modes of highlighting commonalities and variation across retrieved sentences. The writer can hover over a retrieved sentence \textit{(D)} to see more of the context in which it appeared, and save, annotate, and share retrieved sentence examples that they think fulfill a purpose particularly well or poorly.}}
  \label{fig:teaser}
  \Description{A screenshot of the user interface of Corpus Studio showing all elements and design features with annotations.}
\end{teaserfigure}

\maketitle

\input{Sections/introduction}

\input{Sections/related_work}

\input{Sections/concepts}
\input{Sections/system}

\input{Sections/data.tex}
\input{Sections/user_evaluation}

\input{Sections/results}

\input{Sections/discussion}

\input{Sections/conclusion}

\begin{acks}
This material is based upon work supported by the National Science Foundation under Grant No. IIS-2107391. Any opinions, findings, and conclusions or recommendations expressed in this material are those of the author(s) and do not necessarily reflect the views of the National Science Foundation. It was also funded by the Deutsche Forschungsgemeinschaft (DFG, German Research Foundation) -- 525037874 and conducted during the tenure of an Alfred P. Sloan Research Fellowship (Grant No. FG-2023-19960). The authors acknowledge the Alfred P. Sloan Foundation for its support. The Foundation had no role in the design, execution, or interpretation of the research.
\end{acks}

\bibliographystyle{ACM-Reference-Format}
\bibliography{main}

\appendix
\input{Sections/Appendix}

\end{document}

%% file: Sections/abstract_final.tex
Many communities, including the scientific community, develop implicit writing norms. Understanding them is crucial for effective communication with that community. Writers gradually develop an implicit understanding of norms by reading papers and receiving feedback on their writing. However, it is difficult to both externalize this knowledge and apply it to one's own writing. We propose two new writing support concepts that reify document and sentence-level patterns in a given text corpus: (1) an ordered distribution over section titles and (2) given the user's draft and cursor location, many retrieved contextually relevant sentences. Recurring words in the latter are algorithmically highlighted to help users see any emergent norms. Study results (N=16) show that participants revised the structure and content using these concepts, gaining confidence in aligning with or breaking norms after reviewing many examples. These results demonstrate the value of reifying distributions over other authors’ writing choices during the writing process.

%% file: Sections/introduction.tex
\section{Introduction}

Inexperienced researchers often lack knowledge of the implicit writing conventions and structural expectations of particular communities. While existing writing guides offer general advice, they are often too abstract and high-level, making it challenging for junior researchers to apply this guidance in their own work.

Given that people commonly learn through examples, providing access to relevant examples can help bridge this gap. However, there are obstacles: viewing too few examples may lead to overfitting and unintentional plagiarism, while viewing too many examples---without sufficient cognitive supports within the interface---may overwhelm the writer.

Prior work motivated the design of writing support tools with social cognitive theories and cognitive apprenticeship theory~\cite{Hui2023LettersmithSW, Hui2018IntroAssistAT} to help writers learn complex writing skills in a professional setting. While this approach allows novices to learn by mimicking model examples, it also requires expert knowledge to create and annotate these examples, which is often time-consuming and not readily available for many communities.

Learning from examples has also been extensively covered in the language learning literature. Non-native English speakers (NNES) may hesitate to choose from AI-suggested content because they struggle to evaluate the content's fluency and naturalness~\cite{Kim2023TowardsEA, Kim2024HowNE, Ito2023UseOA}. To address this problem, researchers proposed that novices learn by inspecting sentences from a curated corpus such as the Corpus of Contemporary American English (COCA)\footnote{https://www.english-corpora.org -- accessed 12.09.2024} or English dictionaries~\cite{Chang2015WriteAhead2ML} to help students learn the correct use of collocations and phrases~\cite{Mansour2017UsingCT}. To this end, \textit{LangSmith} proposed a writing support tool for drafting academic NLP papers~\cite{Ito2020LangsmithAI} by generating grammatically correct phrases given the user's current draft along with prior section and paper titles. In contrast, our work surfaces existing text from a relevant peer-reviewed text corpus. Furthermore, by avoiding automatically generated text that may---without the writer's knowledge---cross the line into plagiarism if used verbatim, the writer only sees previously published example sentences and their provenance, retaining informed control. 

To this end, we implemented \system, a writing environment with two novel writing support concepts, one at the document level and one at the sentence level, that help writers learn about the writing styles and expectations of particular audiences. These writing supports are designed to expose writers to as many examples as possible to increase the variety of examples shown to the users---with the cognitive support necessary to help them see the emergent patterns that capture otherwise implicit community-specific writing content and style expectations. 
Specifically, at the document level, \system{} computes and renders an \textit{ordered} distribution ~\cite{gero2024supporting} over section title sequences in prior papers from the community the writer wishes to write for. This gives the writer a sense of common and uncommon section titles and in what relative order they appear in prior papers (\cref{fig:teaser}A). At the sentence level, \system{} retrieves many example sentences from the corpus given the user's draft-in-progress and their current cursor location within it (\cref{fig:teaser}C). Words across these many example sentences are algorithmically highlighted or de-emphasized to help users see any emergent norms and outliers. We also allow authors to capture their thoughts and deductions from viewing these examples with bookmarks and notes.

We conducted a user study with \studyN{} participants from the academic community who were actively writing their manuscripts. They used \system{} to draft their paper outline, wrote one to two sections of their manuscript, and answered questionnaires eliciting their overall experience writing within \system.

Our results reveal that users felt more confident in their writing when they discovered that other people in their target community have written in a similar way. Inspecting many examples at once was useful for identifying broad structural patterns, especially for sections that typically have more common structure, such as the description of the system implementation details or participants' demographics.

In summary, we contribute the following:
\begin{itemize}
	\item We designed two novel writing support concepts that make visible emergent patterns at the document-level and sentence-level of a given paper corpus.
	\item We implemented these writing support concepts within \system, an interactive writing environment that helps writers retrieve, inspect, generalize from, and reflect on many examples from their target venue
	\item \changenote{We found that users' writing goals guided their perception of relevance and how they adapted content from the corpus.}
\end{itemize}

%% file: Sections/related_work.tex
\section{Background and Related Work}
In this section we provide a background on writing norms in different communities and how members learn about these norms. We then give an overview of interactive tools that support writing and tools for finding related work.

\subsection{Community Writing Norms}
\input{Sections/related_work/writing_norms}

\subsection{Writing with External Text}
\input{Sections/related_work/writing_support_tools}

\subsection{Finding Relevant Examples}
\input{Sections/related_work/research_tools}

%% file: Sections/related_work/writing_norms.tex
\subsubsection{The Importance of Understanding Writing Norms} 
\citet{Flower1979WriterBasedPA} highlights the difference between \textit{writer-based} and \textit{reader-based} prose, arguing that the former represents the writers' unrefined thoughts while the latter requires their careful \textit{transformation} with the reader goals in mind while ensuring a \textit{shared language} and \textit{shared context}. 
Prior writing literature emphasizes the importance of writing for an audience~\cite{Benharrak2023WriterDefinedAP, Gpferich2009ComprehensibilityAU, Tausch2002SichVA}, but writing for  academic or professional communities can be especially difficult because they have implicit writing expectations and failing to address them may confuse its members or waste their time with ``unfocused and pointless discussions''~\cite{Flower1979WriterBasedPA}. 
\subsubsection{Aspects of Writing Norms}
Writing norms are comprised of many aspects, including the language, content, organization of information, and scope of the written document. Although academic venues may provide a high-level overview of the expected content and style of writing such as encouraging authors to write ``clearly and concisely, avoid jargon, organize the submission to flow logically and smoothly, provide the right level of detail''\footnote{https://chi2025.acm.org/submission-guides/guide-to-a-successful-submission/ -- accessed: 31 January 2025}, it can be challenging for novice writers to apply these suggestions in their own writing, e.g., \textit{What is the expected level of detail for that audience?}
\subsubsection{Learning about a Community's Writing Norms}
To support writing in an academic context, prior research has proposed the \textit{cognitive apprenticeship} theoretic approach to teach novices complex cognitive tasks by mimicking experienced members of the community \cite{Collins1988CognitiveAT}. Drawing from these ideas, our work instantiates two strategies of cognitive apprenticeship, namely modeling and reflection.

%% file: Sections/related_work/writing_support_tools.tex
\subsubsection{Writing with \changenote{Retrieved or Generated Additional Text}}
Many writing support tools facilitate writing by suggesting content that users did not write themselves. Sentence completions are a frequent form of writing assistance~\cite{Bhat2023suggestioninteraction, Lee2022CoAuthorDA}. However, direct content generation can have negative impacts such as diminishing the author's self-expression and ownership over the final text outcome~\cite{Li2024TheVB}.
Related work has also looked into providing writing inspirations~\cite{Singh2023elephant, Gero2021SparksIF, Chung2022TaleBrushSS}. 

\changenote{We build on this line of research by surfacing prior text as the user writes new text (e.g., \citet{Naeem2018ASE} and \textit{IntroAssist}~\cite{Hui2018IntroAssistAT}). A critical difference is the much larger scale of prior text surfaced to the user and post-processed for emergent patterns in our work.} For example, \textit{IntroAssist}~\cite{Hui2018IntroAssistAT} presents expert-curated email examples to help users draft their help-seeking message in a professional context, while 
our approach uses computational methods to show many examples drawn from a large text corpus of previously published papers.

\subsubsection{Writing with Text from a Corpus}
\citet{Hui2023LettersmithSW} proposed a writing support tool to help college students write in a professional context assisted by \textit{scaffolded annotations}. The authors proposed a \textit{checklist} with multiple annotations that provide cues about the content to include in the professional email. This checklist is complemented with \textit{model texts} where sections of that text have been annotated with the elements of the checklist. This approach helps users understand the overall structure and expected content while also demonstrating concrete instantiations of the high-level guidelines. Our concepts 
help users plan the structure and content of their writing by showing many concrete examples of how other members of their target community have written similar texts.

%% file: Sections/related_work/research_tools.tex
As previously described, writers learn from inspecting examples, but finding related work with the purpose of writing support is challenging. Search engines such as Google Scholar, the ACM Digital Library, Mendeley, and Semantic Scholar typically surface documents based on topical similarity while writers need specific analogous snippets of text from the manuscript to inform their writing. Current tools also help researchers understand how their own work relates to other work~\cite{paperweaver_lee2024} by contextualizing embedded citations~\cite{chang2023citesee}, drawing connections to follow-on work directly in the paper document~\cite{rachatasumrit2022citeread}, and making recommendations based on a paper's metadata such as citations and authorship~\cite{Cohan2020SPECTERDR, Singh2022SciRepEvalAM, Kang2022AugmentingSC, Portenoy2021BurstingSF} or its semantic similarity~\cite{Hope2017AcceleratingIT, Kang2022AugmentingSC, Chan2018SOLVENT}.
In contrast, our design focuses on finding related work for the purpose of writing support, i.e., that helps writers identify existing writing conventions from a specific community.

Most interactive research tools either support analyzing related work \cite{Ponsard2016PaperQuestAV, paperweaver_lee2024, Mackinlay1995AnOU, Lee2005UnderstandingRT} or writing with computational help \cite{Lee2022CoAuthorDA, Dang2022beyondTextGen}. 
In existing contexts, writers often need to navigate away from their writing environment to search for related work, scroll through the target document, select and process relevant text structures and draw conclusions about the implicit writing conventions encoded in the target document. Performing all of these steps leads to abrupt context switches and breaks the writing flow. Our approach proposes an integrated environment to support writing and searching for model text in the same interface, thereby minimizing the number of context switches and streamlining the writing and research process.

%% file: Sections/system.tex
\section{Corpus Studio}
\input{Sections/system/corpus_studio_intro}

\input{Sections/system/section_titles}
\subsubsection{User Notes}\label{sec:user_notes_view}
\input{Sections/system/user_notes}

\subsubsection{Contextual Tooltips}\label{sec:tooltips}
\input{Sections/system/tooltip}

\subsubsection{Features for Promoting Learning, not Plagiarism}\label{sec:plagdesign}

\changenote{A combination of features are intended to work together to support users in actively engaging  with many examples, rather than anchoring on and copying any one. More specifically, showing multiple examples at once may help novice writers make more informed decisions about plagiarism: when sentences are highly similar across papers, this often indicates accepted community writing norms (like in methodology sections), while sections with more varied writing styles suggest areas where similar phrasing should be avoided. In concert with showing many examples, two additional design choices are intended to further facilitate learning over plagiarism: First, we disable the user's ability to copy and paste retrieved sentences into their own draft. Second, by providing visible affordances for saving and annotating their insights, we may promote deeper reflection on the writing patterns they observe.}

\subsection{Implementation Details}
\label{sec:imp}
\input{Sections/system/implementation}

%% file: Sections/system/corpus_studio_intro.tex
We designed \system{} to support writers in learning about and applying community-specific writing norms. This is achieved primarily through novel mechanisms for the retrieval and rendering of content from a corpus of existing documents written by and for that community.

\subsection{Design Goals}

\changenote{Our overarching design goal is to \textit{empower users with evidence of community-specific emergent norms and outliers} at the document and sentence level as they write their own document for the same audience.} 
\changenoteAcc{In the process of fulfilling this overarching goal, our goal is to reduce, not facilitate, unintentional human- or AI-generated plagiarism.}

\subsection{Usage Scenario}
\input{Sections/system/usage_scenario}

\subsection{System Characteristics}

\changenote{In this work, the evidence of community-specific emergent norms and outliers at both the document and sentence level is provided to writers in the form of \textit{extracted examples} from a corpus of documents written for and by their intended audience.} 
\changenoteAcc{To facilitate the writer's recognition of these emergent norms from examples without facilitating the writer's anchoring on any one example too much, we look to theories of human concept learning:}

\changenote{\paragraph{Variation Theory} Regardless of whether a human or a machine is learning from examples, any learning process can fall victim to overfitting, i.e., learn a rule that does not generalize to the rest of the corpus, when trained on too little data. For example, the Variation Theory of Learning~\cite{Lo2012VariationTA} was explicitly designed to minimize human overfitting when learning from examples. With only a few examples or suggestions, previous systems may cause users to anchor~\cite{tversky1974judgment} on just one of them. By instead showing \textit{many} examples, we hope the user may be (1) less likely to anchor to one of the retrieved examples (or even unintentionally plagiarize from it) and (2) more likely to observe community writing norms and possibilities without overfitting to the retrieved examples.}

\changenote{\paragraph{Structural Alignment Theory} While the many retrieved examples in our work are not algorithmically selected on the basis of Variation Theory, the textual examples are ordered and rendered with visual attributes (e.g., grayed-out repeated words or color-coordinated highlights) designed to support the recognition of emergent commonalities and outliers, based on the human cognitive processes of pattern recognition described by Structural Alignment Theory~\cite{gentner2017analogy}.}

\subsubsection{Text Editor}
\input{Sections/system/texteditor} 
\input{Sections/system/creating_section_headers}

\subsubsection{\changenoteAcc{Surfacing Document-level Community-specific Emergent Norms and Outliers}} \label{sec:doclevel}

Inspecting outlines from other papers in the target community can help writers develop their own. 
\changenote{To help writers understand how papers are commonly (and uncommonly) structured within their target community at the document level, we show an ordered distribution of section title clusters from the entire corpus} \changenoteAcc{immediately to the left of the central text editor (\Cref{fig:teaser}A \emph{Section Titles From Other Papers}). This ordered distribution is computed by applying Positional Diction Clustering (PDC)~\cite{gero2024supporting} to the ordered lists of top-level section titles extracted from each paper in the corpus.} 

\changenoteAcc{PDC is a structural mapping engine~\cite{falkenhainer1989structure} designed to algorithmically emulate at scale the analogical matching process humans do when comparing two or more structurally similar objects, as described by Structural Alignment Theory~\cite{gentner2017analogy}.} 
In \system{}, the objects are papers, and the resulting order of section title clusters that PDC generates preserves the section titles' approximate relative positions in their respective papers. \changenoteAcc{For example, in \Cref{fig:teaser}A, \texttt{Introduction} is the ubiquitous starting point\footnote{Note that the screenshot in \cref{fig:teaser} was taken with the system running on a subsample of the corpus, explaining why there are e.g. only 128 introductions, extracted from 128 papers.} %
	and, in approximately 80\% of papers, it is followed by \texttt{Related Work}. The user can also see outliers like the two papers that have a section titled \texttt{An Interactive Walkthrough} soon after.} 

Section title clusters are shown with a count and a proportionally long bar indicating the total number of section titles in that cluster found across the entire corpus. The most frequent section title in each cluster is chosen as the cluster name. %
Clicking on a cluster name expands (or closes) a cluster, showing (or hiding) all its members.

We also provide cognitive support for perceiving the commonalities and variations 
within each cluster of section titles:  
The underlined words within each cluster name appear in every section title within it (a strict commonality), serving as information scent about the cluster contents when it is closed. %
Within each cluster, %
the PDC algorithm dictates the order of section titles to minimize diction differences between section titles in adjacent rows;
words that appeared previously in the ordered list of cluster members are greyed out so that variations might be more visible. %
This text graying strategy is based on the interleaved view of PDC results in \citet{gero2024supporting}. %

\subsubsection{\changenoteAcc{Surfacing Sentence-level Community-specific Emergent Norms and Outliers}} \label{sec:sentencelevel}

This writing support is intended to provide insights into both established conventions and areas where variation is common at the sentence level.
By retrieving and displaying many relevant sentences simultaneously, users can observe how similar content is expressed across different papers. These distributions can reveal both rigid conventions (narrow distributions) and areas of flexibility (wide distributions). For instance, participant descriptions in user studies often follow stereotypical patterns, while discussion sections typically show more variation to accommodate novel findings.

\changenote{\paragraph{Spatial Retrieval} At the core of our approach is a novel retrieval mechanism we call \textit{analogous spatial retrieval}, 
	which integrates context, content, and spatial information. Users can request this retrieval of sentences from other papers by hitting the \texttt{TAB} key
	as they are drafting their text.
	Given 
	where the user's cursor is placed in the text editor at the moment of retrieval (i.e., the text of the currently focused
	paragraph cell and the title of the section containing it), 
	we retrieve and display many potentially analogous sentences from similarly titled sections of papers in the corpus
	(\cref{fig:teaser}C \textit{What others wrote} column}) as well as the subsequent sentence in each paper (\cref{fig:teaser}C \textit{What they wrote next} column). See \cref{sec:imp} and \cref{sec:emb} for more spatial retrieval implementation details. %

\input{Sections/system/offset_querying}

\paragraph{\changenoteAcc{Scale of Retrieval}} \changenoteAcc{Following Gentner’s work on analogical learning theory ~\cite{gentner2017analogy}, the comparison of many example sentences promotes the extraction of common principles or abstract schemas, which may be more likely to be transferred to future writing tasks.} We experimented with displaying 10, 25 and 50 sentences at once and found that 25 sentences provided enough content for writers to explore \changenoteAcc{and potentially recognize common principles and abstract schemas} while not overwhelming them.

\paragraph{\changenoteAcc{Sentence Rendering.}} \changenoteAcc{However, what writers learn from these examples depends not only on their scale and variety, but also on how the writer mentally processes and compares them~\cite{gentner2017analogy}.} To aid in this mental sentence comparison process, the sentence examples can be rendered with algorithmically assigned visual variables (e.g., color-coordinated highlights) designed to support the recognition of emergent commonalities and outliers (and possibly support the mental cross-sentence alignment process necessary for developing abstract sentence schemas).
\cref{fig:highlighting} shows a comparison of the two available but optional rendering modes: %

\begin{itemize}
	\item \textit{Distinctly Colorize Recurring Words.} We created a color map based on the twenty most-frequently occurring words across all retrieved sentences and columns (excluding stop-words), then iterated through each retrieved sentence and colorized the other word occurrences using the color map. This mode is the default.
	\item \textit{Grey Out Repeated Words.} This mode de-emphasizes previously seen words by iterating through each retrieved sentence and greying out words that have appeared in any of the previous sentences in the result list. 
\end{itemize}

Re-ranking is another way for users to explore relationships across sentences. 
When users click on the re-ranking icon (arrow below bookmark icon next to each sentence), the selected sentence moves to the top of its column in \cref{fig:teaser}C. \system{} then sorts the other sentences in ascending cosine-distance from that new top sentence.
This triggers the system to recompute the greyed-out words to account for the new sorted column.

\

\input{Sections/system/displaying_retrieved_sentences}

%% file: Sections/system/usage_scenario.tex
Alice has been working on a new interactive writing support system and is drafting her manuscript for submission to an HCI venue. She has some prior experience writing academic papers but is unfamiliar with writing papers with a focus on a system contribution. She opens \system{} to learn more about her target venue's writing expectations while working on her draft. %
\paragraph{Inspecting the section titles} She looks at the ordered distribution over section titles.  %
As shown in \cref{fig:teaser}A, she expands the cluster named \textit{Implementation}, which typically comes after \textit{System Design} and around the same location as \textit{Methods} (when those sections exist in a given paper). In the expanded cluster, she observes how other authors have phrased this typical section's title, i.e., more commonly \textit{Implementation} and less commonly \textit{Implementation Details} or \textit{Implementation of <System Name>}. Although she likes the idea of adding a system name to the title, she decides to stick with the more generic commonly chosen option for now, and finishes her paper outline of section titles in the text editor beside it. %
\paragraph{Inspecting the retrieved sentences} In the text editor, she starts drafting the implementation section immediately under the \textit{Implementation} section title, but she does not know how she should introduce it. She writes a few words and, without moving the cursor, hits the \texttt{TAB} key to retrieve related sentences from other work to get a sense of what (and how) other people wrote the beginnings of their implementation sections. She notices that many authors start with \textit{``We implemented} \texttt{<name of the system>} \textit{to do/test} \texttt{<some hypothesis>}\textit{''} while some also state their design rationale. %

\paragraph{Developing a personal style sheet} She bookmarks one of the retrieved sentences to remind herself of this observation and adds the note: ``The first paragraph of the implementation is often formulaic and with this recognizable structure.'' She concludes that it is probably well accepted in the community if she starts her paragraph in a similar way. %
\paragraph{Continuing Writing} Alice continues to write down her ideas in the text editor and occasionally retrieves example sentences given where her cursor is when she needs inspiration or wants to confirm how other authors have written their manuscripts.

%% file: Sections/system/texteditor.tex
Users outline their paper and draft their prose in the central text editor (\Cref{fig:teaser}B). The editor supports two text cell types: section title and paragraph. 

%% file: Sections/system/creating_section_headers.tex
Borrowing from the syntax of Markdown editors, the editor allows users to create section titles by pre-pending text with \#'s where the number of \#'s indicate the level of the section title.

%% file: Sections/system/offset_querying.tex
To further help writers ``peek into the future,'' 
that is, explore what other authors wrote later in a given section, users can insert empty paragraph cells after a non-empty paragraph cell and then trigger sentence retrieval. The number of empty cells corresponds to the offset \texttt{o} (in sentences) to be retrieved from the corpus. 
Retrieving content with an offset works by first retrieving all sentences that would have been retrieved if the cursor had been in the nearest preceding non-empty paragraph cell but then retrieving and displaying the \texttt{o}-th subsequent sentences.

%% file: Sections/system/displaying_retrieved_sentences.tex
\begin{figure}
	\centering
	\includegraphics[width=\minof{\columnwidth}{0.5\textwidth}]{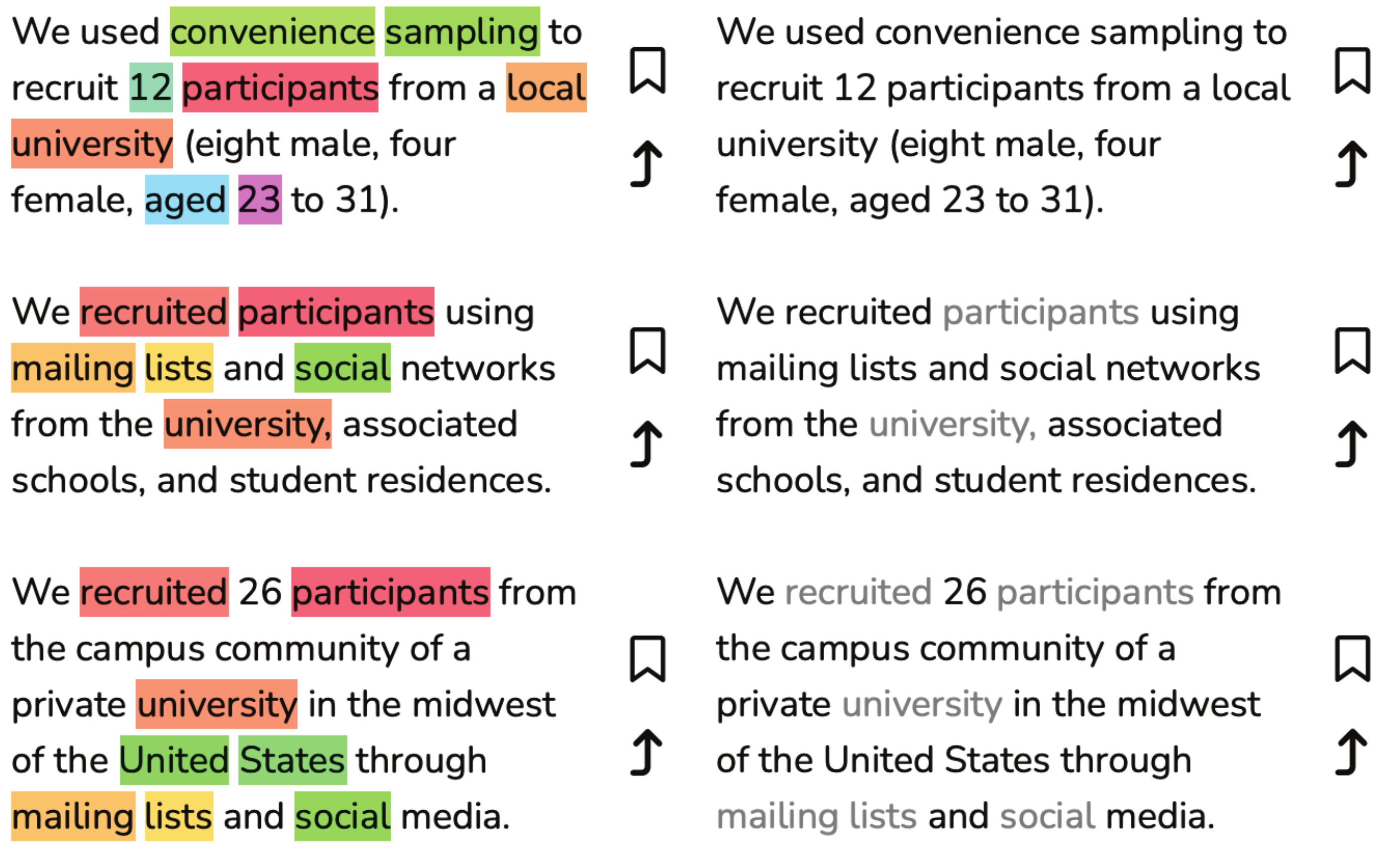}
	\caption{\changenote{Side-by-side comparison of sentence rendering modes to help the writer identify textual commonalities and variations across retrieved sentences: (left) highlight each commonly recurring word with the same distinct color versus (right) greying out repetitions of words in subsequent sentences}}
	\Description{A screenshot showing distinctly colorized and grey out recurring words side by side}
	\label{fig:highlighting}
\end{figure}

%% file: Sections/system/user_notes.tex
\begin{figure}
	\centering
	\includegraphics[width=\minof{\columnwidth}{0.5\textwidth}]{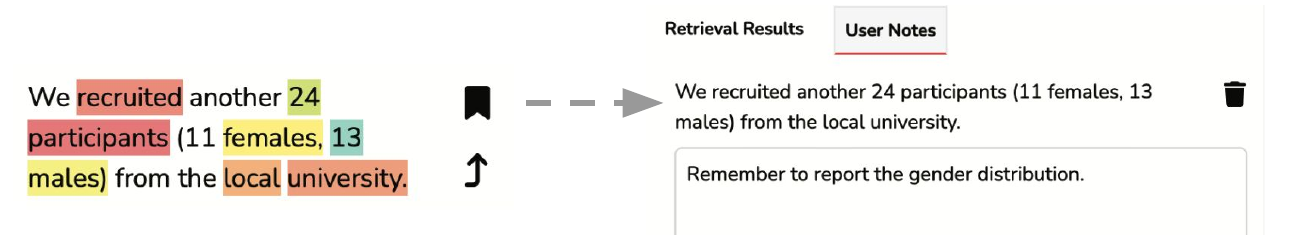}
	\caption{\changenote{Illustration of the bookmarking feature showing how writers can save sentences and add notes, with the original bookmarked text preserved for context. In this example, the writer added a reminder to note the gender distribution of participants.}}
	\Description{A screenshot showing a retrieved sentenced that has been bookmarked, indicated by the filled bookmark icon. Next to the bookmarked sentence there is a screenshot of a user note with some notes that relate to the previously bookmarked sentence.}
	\label{fig:bookmark}
\end{figure}

As shown in \cref{fig:bookmark} and the top right callout in \cref{fig:teaser}C, users can bookmark a sentence and then comment on it in the User Notes section, e.g., with a reminder of why it fulfills its purpose particularly well or poorly. 
These annotated sentence examples can be exported as a \texttt{csv} file to facilitate sharing observations, preferences, and insights with others.

%% file: Sections/system/tooltip.tex
Writers may want to know more of the context surrounding a retrieved sentence, e.g., to avoid misinterpreting it. 
As shown in \cref{fig:teaser}D, users can hover over a retrieved sentence to view a floating window showing the paper's title, the hierarchical section title path, and the sentences immediately before and after.
The paper title links to the full paper in the ACM Digital Library. 
Additionally, if any retrieved sentence includes a citation, users can hover over it to display the full citation information, including the authors and the referenced paper title.

%% file: Sections/system/implementation.tex
\changenote{
	We implemented our system as a web application using SvelteKit.\footnote{https://kit.svelte.dev} For sentence retrieval, we used WeaviateDB\footnote{https://weaviate.io} as our vector database, using its built-in text embedding capabilities and approximate k-Nearest Neighbor (kNN) search implementation \cite{malkov2018efficient}. The system was deployed on a workstation with 12 GB of GPU memory and made remotely accessible to study participants.
}

\changenote{
	While determining relevance in retrieval systems is inherently complex and depends on users' dynamic writing objectives, querying based on semantic similarity in text vector space is an established approach in retrieval-based systems \cite{memolet_yen2024}. Our initial implementation explored hybrid search techniques, following prior work that used Reciprocal Rank Fusion \cite{Rackauckas_2024} to combine keyword-based and vector-based retrieval results. However, through initial experiments, we found pure vector-based search provided more contextually relevant matches. 
}

%% file: Sections/data.tex
\begin{figure*}
	\centering
	\includegraphics[width=\linewidth]{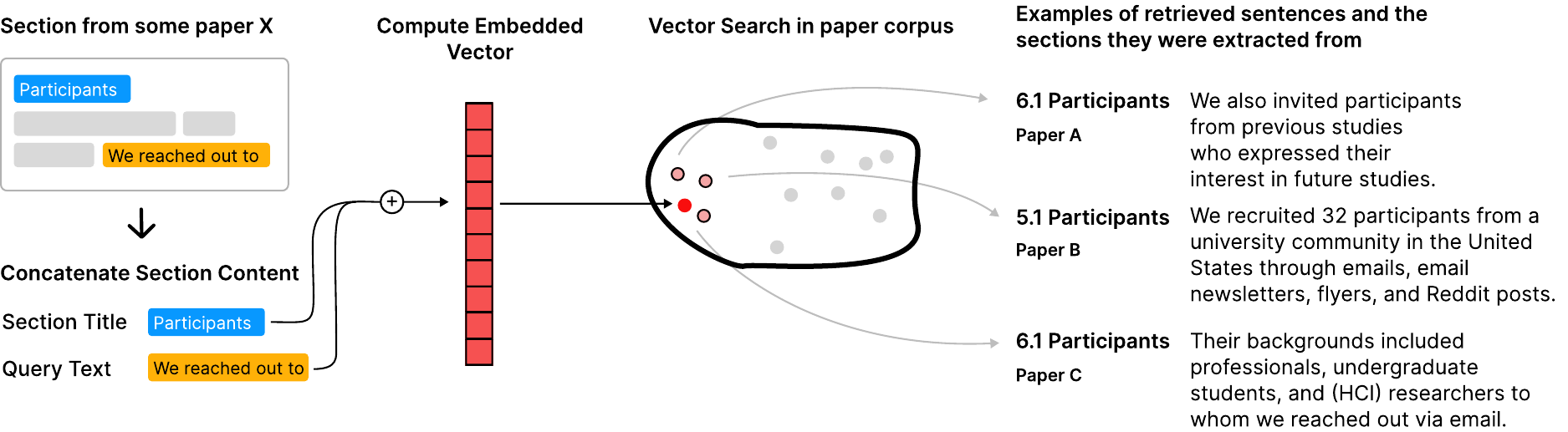}
	\caption{\changenote{Sentence embedding workflow designed to retrieve analogous content across papers by incorporating section context. The process prepends section titles to sentences before vectorization, enabling the system to find similar content from matching sections (e.g., `Participants' sections) across different papers, as shown in the three retrieved examples on the right.}}
	\Description{An illustration of the embedding process.}
	\label{fig:embedding-process}
\end{figure*}

\section{Data \& Processing}

We collected 478 papers from two HCI venues (IUI and UIST) which included the proceedings from 2021 until 2023. These venues typically publish HCI system contributions. We completed the data request process with the ACM.

\subsection{Pre-processing Documents} We initially used the NLTK module to split each manuscript into sentences, but we experienced many edge cases where extracted sentences were incomplete. We therefore chose an LLM-based approach instructing the model to separate sentences by inserting \texttt{[EOS]} tokens between sentences.
For each extracted sentence, metadata was also saved, e.g., the context shown in the tooltips described in \cref{sec:tooltips}.

\subsection{Embedding Process} \label{sec:emb}
To implement spatial retrieval, we follow an approach resembling \textit{LangSmith}~\cite{Ito2020LangsmithAI}. We prepared our corpus to support this functionality by pre-pending section titles to the extracted sentences before embedding the concatenated sequence using OpenAI's `text-embedding-small-3` model \changenote{(\cref{fig:embedding-process})}. This approach enables \system~ to retrieve related sentences based on both the section title and the content of the sentence. 
We used default parameter values and default un-trimmed vector dimensions.

\subsection{Extracting an Ordered Distribution over Section Titles}\label{sec:pdc_titles}
The sequences of top-level section titles for each paper were extracted and organized into a list of lists using positional diction clustering (PDC)~\cite{gero2024supporting}. This is identical to the process that produced the data for the Interleaved PDC view in \citet{gero2024supporting}, except instead of processing LLM responses as sequences of sentences, we processed papers as sequences of section titles. We also computed how many papers had each section title in order to populate a corresponding histogram in the \system~ interface.

We experimented with different sample sizes ranging from 200, 500, 1000, to 5000 section titles and found that distinctive distribution characteristics quickly emerged even with smaller sample sizes. 
For our prototype we decided to render and cluster 1000 section titles to demonstrate both the defining characteristics of the section title distribution while still providing a variety of ways authors have organized and phrased their section titles.

%% file: Sections/user_evaluation.tex
\section{User Study}
To evaluate our two novel writing support concepts (described at the document level in \cref{sec:doclevel} and at the sentence level in \cref{sec:sentencelevel}), we used \system{} to conduct a controlled within-subjects experiment with participants who were, with one exception, actively working towards writing a manuscript for an HCI venue.
Participants completed two writing tasks during the user study: (i) drafting or revising an outline of their paper and (ii) drafting or revising section(s) of their paper. We referred to \system~ as a ``set of design features'' or ``prototype'' during our user evaluation of the concepts. 
We used Google Docs as our baseline interface because it is a commonly used writing environment in this context. 
We aimed to investigate how participants would use the writing supports to achieve their goals, how the supports would affect their writing process, and how they may help them identify norms of a target community.

\aptLtoX[graphic=no,type=html]{\begin{table}[h]
\centering
\setlength{\tabcolsep}{4pt}
\begin{tabular}{lp{14pc}p{6pc}}
\toprule
\textbf{Category} & \textbf{Details} &  \\ 
\midrule
Participants & {Total: 16 (8 men, 8 women)} \\
& {Age: 18-24 (4 participants), 25-34 (11), 35-44 (1)} \\
& {Criteria: Over 18 years old, fluent in English, actively writing a manuscript for an HCI venue} \\
\multirow{2}{2cm}{Writing experience} & General writing & 4.3 years (SD 2.5) \\
& For target venue  & 2.6 years (SD 2.8)\\
\multirow{2}{2cm}{Educational background} & Undergrad. students & 3 \\
& Graduate students         & 6 \\
& Post-doc & 3 \\
& Faculty members           & 3 \\
& Research associate        & 1 \\
Target venue  & CHI & 11  \\
& CHI or NeurIPS            & 1 \\
& NeurIPS                   & 1 \\
& IUI                       & 2 \\
& Undecided (HCI work)      & 1 \\ 
\bottomrule
\end{tabular}
\caption{Participant information}
\label{tab:participant_info}
\end{table}}{\begin{table}[]
\centering
\newcolumntype{L}{>{\raggedright\arraybackslash}X}
\newcolumntype{P}[1]{>{\raggedright\arraybackslash}p{#1}}
\renewcommand{\arraystretch}{1.2}
\setlength{\tabcolsep}{4pt}
\begin{tabularx}{\linewidth}{P{2cm}P{3cm}L}
\toprule
\textbf{Category} & \textbf{Details} &  \\ 
\midrule
Participants & \multicolumn{2}{l}{Total: 16 (8 men, 8 women)} \\
& \multicolumn{2}{l}{Age: 18-24 (4 participants), 25-34 (11), 35-44 (1)} \\
& \multicolumn{2}{P{6cm}}{Criteria: Over 18 years old, fluent in English, actively writing a manuscript for an HCI venue} \bigskip\\
\multirow{2}{2cm}{Writing experience} & General writing & 4.3 years (SD 2.5) \\
& For target venue  & 2.6 years (SD 2.8)\bigskip\\
\multirow{2}{2cm}{Educational background} & Undergrad. students & 3 \\
& Graduate students         & 6 \\
& Post-doc & 3 \\
& Faculty members           & 3 \\
& Research associate        & 1 \bigskip\\
Target venue  & CHI & 11  \\
& CHI or NeurIPS            & 1 \\
& NeurIPS                   & 1 \\
& IUI                       & 2 \\
& Undecided (HCI work)      & 1 \\ 
\bottomrule
\end{tabularx}
\caption{Participant information}
\label{tab:participant_info}
\end{table}}

\subsection{Participants}
Sixteen participants were recruited through mailing lists and Slack channels from different universities in the USA, Germany, United Kingdom and South Korea  (\cref{tab:participant_info}). Participants had most commonly authored or co-authored 3-5 academic papers already, but two had authored less and four had authored more than twenty (see Appendix \cref{fig:papers_written} for the entire distribution). The average Need for Cognition (NCS-6) score~\cite{ncs6} across all participants was high---4.57 out of 5 (standard deviation: 0.46)---so participants were particularly willing to exert cognitive effort on a given task.

Most participants were targeting HCI venues, and one was targeting a machine learning venue. All participants reported referencing published papers during their writing process: 50\% always did, 30\% (5/16) often did, and 20\% (3/16) sometimes did.

All but one participant had previously used at least one writing support tool. Tools included Grammarly (12), DeepL (3), ChatGPT (4), Google Docs add-ons for reference search (1), Zotero plugins (1), SciSpace (1), and OpenAI’s Grammar Checker plug-in (1). Usage frequency varied: 2/16 used them occasionally, 6/16 used them a few times per month, 5/16 a few times a week, and 3/16 once a day or more.

\subsection{Study Procedure}
All studies were conducted remotely via Zoom and were evenly facilitated between the two lead co-authors. \system{} was accessible as a web application, so participants were not required to install anything. Each study took approximately 60-75 minutes, and participants received \$25 (USD) via digital payment (Zelle, Venmo, or PayPal) as compensation for their time.

\subsubsection{Consent \& Pre-Study Survey} Each participant, after giving informed consent, completed a brief survey about their demographics, experience with scientific communication and writing for a target venue, current strategies for writing, use and experience with writing support tools, and questions relevant to the task that they would be completing. All survey questions are included in \cref{app: surveys}.

\subsubsection{Task Preparation}
Participants were informed that \system's corpus, from which they would be viewing section titles and retrieved sentences, consisted of papers from two HCI venues, IUI and UIST, between 2021 and 2023. This information was disclosed to ensure participants had a clear understanding of the corpus context, allowing them to engage with the examples accordingly, regardless of their own specific target venue.

Before starting each of the two tasks, participants watched a tutorial video that explained the relevant writing support concept and its supporting features in \system. Participants were told that we evaluated the features, not their work.

\subsubsection{Task 1: Outline Writing} \label{sec:task_1}
\changenote{We chose outlining as our first task because it let us examine how participants structured their work after seeing an ordered distribution over section titles from published papers.}

\changenote{The outlining task had two sub-tasks: drafting section titles for the first and second half of the paper. Participants used either a baseline interface (Google Docs) or the exploratory interface (\system) in counterbalanced order. With the baseline, participants relied on their typical resources. With \system, they just used the \cref{fig:teaser}A \emph{Section Titles From Other Papers} feature. Each sub-task had a 5-minute limit. They thought aloud while writing and completed post-task surveys (\cref{app:t1-baseline}, \cref{app:t1-exploratory}).}

\subsubsection{Task 2: Paper Writing}  \label{sec:task_2}	
Participants began writing or revising sections of their 
manuscript for their target venue. They were given the option to bring the outline created in the previous task and any supporting text that they may have already written into this task. Participants were allowed to use any feature of \system{} to help them write. As part of the task, participants were asked to bookmark example sentences that they would want to save and annotate in User Notes for themselves or a mentee. They were given twenty minutes to complete the writing portion and five minutes at the end to revise and update their User Notes; \csnote{they also were asked to think aloud while completing this task.}

\csnote{Note that, in the tutorial video at the beginning of the task, we included the information about plagiarism and patterns over retrieved sentences described in \cref{sec:plagdesign}. We also provided the following instruction: ``When there is high diversity of sentence content and style, read many sentences before composing your thoughts to avoid mirroring any one sentence directly; this will hopefully help you avoid plagiarism and ensure your work remains authentic.''}

\subsubsection{Post-Tasks Survey and Interview}
To conclude the task and study, participants completed a post-task survey (see \cref{app:t2-survey}) followed by a brief semi-structured interview (see \cref{app:interview}).

%% file: Sections/results.tex
\section{Qualitative Results }
The two lead co-authors analyzed the user study transcripts with a \textit{Grounded Theory} approach~\cite{Jordan1994BasicsOQ}. They separately collected a list of participants' behaviors, approaches, insights, and suggestions about the interface as well as responses to initial open-ended survey questions and the final semi-structured interview questions. To derive high-level themes, both co-authors jointly organized and clustered the codes through affinity diagramming. Throughout and after this process, co-authors discussed their respective codes, resolved assignment conflicts through discussions, and refined the labeling of the clusters. Once both authors agreed on the clusters, they distributed the themes evenly among the authors who individually identified further evidence for each theme in the study transcriptions. Lastly, all authors reviewed and agreed on the final result report.

\paragraph{Overview.}
\csnote{Participants reported that seeing retrieved sentences and section titles helped them overcome the blank page problem and continue drafting. They also found it helpful to view multiple contextually relevant sentences from various papers on a single page within a centralized writing environment, which allowed for easier comparison of structure and patterns. They reported that this layout was more efficient and less overwhelming than traditional methods, such as having multiple papers open to compare texts.} 

\csnote{Participants used the retrieved sentences and section titles to gather ideas or mimic good writing, and found the tool helpful for tasks such as drafting introductions, starting sections, describing systems, and beginning paragraphs.} 

\csnote{Some participants reported that their understanding of a venue's norms remained the same after utilizing the tool, but they still found value from other aspects such as terminology, phrasing, structure, and ideas. For example, a participant noted that they were inspired to think more critically about how they wrote rather than just following norms.}

\begin{figure*}
\centering
\includegraphics[width=0.6\textwidth]{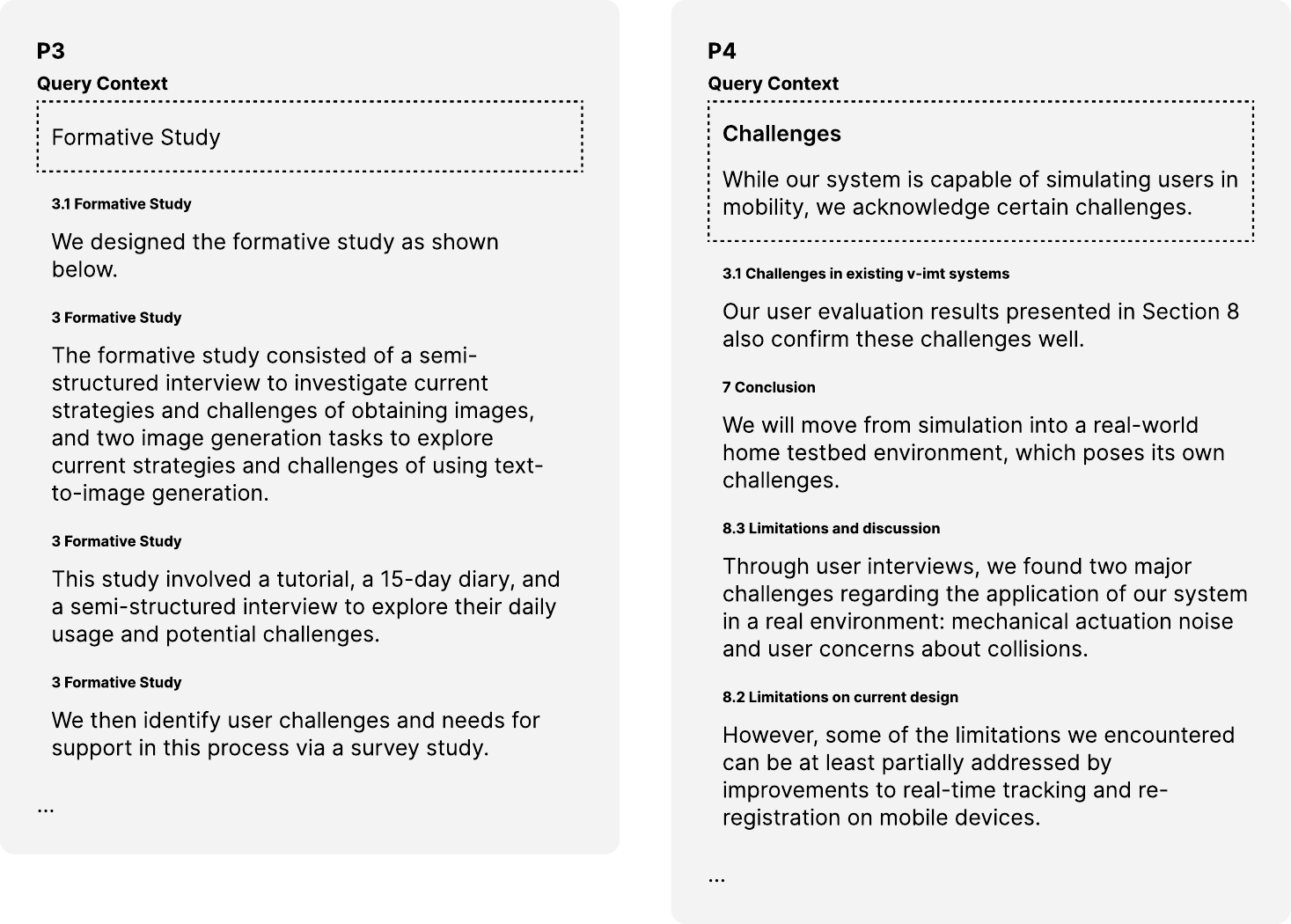}
\caption{\changenote{Examples of different users' immediate writing contexts when requesting sentence examples, and the first of \system's retrieval results. Note that these are representations of system logs, not user interface screenshots.  (Left) P3 
		queried for inspiration on how to introduce a formative study section, retrieving examples of how other papers structured and presented their study introductions. (Right) P4
		searched for ways to acknowledge design and study challenges, finding various examples of how other papers discussed limitations and technical constraints. }
}
\Description{A side-by-side example of retrieved sentences for P3 and P4}
\label{fig:user_phrases}
\end{figure*}

\subsection{\changenoteAcc{Document-level Writing Support}}
\label{sec:utility of section titles}
\input{Sections/results/approach_section_titles_from_other}

\subsection{\changenoteAcc{Sentence-level Writing Support}}

\input{Sections/results/writing_strategies_with_retrieved_sentences}

\subsubsection{\csnote{Sentence Rendering Modes}}
\label{sec:visual_support}

\changenote{
We analyzed participants' perception of the cross-sentence structure mapping engine support \csnote{features and how these features helped participants process the retrieved sentences.} 
}
\changenote{While initially overwhelming for some, participants found \emph{distinctly colorized recurring words} valuable.
It helped participants scan text efficiently.
\p{7} explained, \userquote{``I can just look [for the] participants in that color rather than spending my whole time reading over every block.''} 
\p{4} noted, \userquote{``I do like the color version the most while it could be perceived as a bit overwhelming... the color does help me find the relevant words.''} \csnote{\p{6} and \p{9} emphasized the ability to read and skim through the text faster.}} \csnote{The feature served multiple purposes: \p{8} found it helpful for providing structure, while \p{13} used it to guide their attention to content (\userquote{``I just looked at where's like the least or the lowest redundancy''}).}

\changenote{
There were more mixed comments on and less use of \emph{greying out repeated words}. \p{8} found it counterproductive: \userquote{``I can't think of a reason that I would want to have those grayed out... These were sort of the key pieces of the sentence that I was trying to match.''} \p{12} did not even notice the feature. In contrast, \p{4} and \p{3} preferred it. \p{3} reported that their eyes were still drawn to greyed-out words, suggesting the words' relevance mattered more than their visual appearance. 
}

\subsection{ \csnote{Bookmark and User Notes}}
\changenote{
Participants used bookmarking and note-taking to manage content during their writing process (\p{1, 4, 10, 11, 16}). Initially, they bookmarked content that \userquote{``caught their interest''} (\p{10}), focusing on topic relevance (\p{4, 8, 10, 16}) and writing style (\p{3, 4}).
When reviewing these bookmarked sentences, participants would either enhance them with notes or unbookmark them if they seemed no longer sufficiently relevant.
}

\changenote{ 
Participants used notes on bookmarked sentences in several ways. 
Some analyzed writing techniques, with \p{11} examining content focus and \p{4} studying structural elements. Notes also served as a teaching tool (\p{6, 8}), with \p{6} particularly valuing sentence-specific annotations for student guidance: \userquote{``If I were planning to share this with someone else I think that'd be really useful, especially to my students [...] Being able to specifically reference the part of that paper, I think is useful.''} However, not all participants found notes helpful---\p{7} preferred editing drafts directly.
}
\changenote{
Overall, this dual system of bookmarks and notes supported both content curation and reflection, while potentially enabling knowledge transfer to more junior community members.
}

\subsection{\csnote{Tooltips}}\label{sec:tooltip}
\input{Sections/results/results_tooltip}

\subsection{\csnote{Use of Features, Especially Across Writing Stages}} 
\input{Sections/results/writing_stages_utility}

\section{Quantitative Results}\label{sec:quant_results}
We analyzed participants' responses to survey questions using a non-parametric Mann-Whitney U test. For Task 1, i.e., outline writing, we analyzed participants' responses comparing the two sub-tasks (first half of the outline vs. second half of the outline) and two interface conditions (using a regular text editor without document-level support vs. with section title rendering in \system). 
For Task 2, the prose writing task, we evaluated participants' reported frequency of using \system~ features and their perceived usefulness.

\input{Sections/results/quantitative_results/task_outlining}

\input{Sections/results/quantitative_results/task_writing}

%% file: Sections/results/approach_section_titles_from_other.tex
\subsubsection{Existing Practices for Creating Section Titles}
\csnote{Prior to Task 1, participants described their methods for drafting outlines. Many saved a list of papers in Google Docs or a local folder, reviewed the individual works, and aligned their section titles with papers they deemed most relevant to their topic. Some participants followed section titles from their own prior work, while others consulted colleagues familiar with their target venue or reviewed papers from the venue to follow a similar structure.} %

\subsubsection{Ways of Reading and Using the Section Titles From Other Papers Feature}
\csnote{
Some participants entered the study with an existing notion of the section titles they needed and used the distribution of section titles as a checklist to validate their choices (\p{1, 3}). \p{13} found the feature useful for identifying structure, particularly for new writers.
Several participants either chose a section title from the distribution and adapted it to align with their writing goals or modified their pre-existing section title based on inspiration from the distribution. \p{7} decided to use the feature to find top-level section titles and then create their own subsection titles. }

\csnote{Participants used section title frequency as an indicator of norm and popularity. \p{10} explained, \userquote{``I'm less drawn to section [titles] that only have two or one [paper they show up in] because it doesn't make me feel confident that it's common.''} }

\csnote{Post-task, participants described scenarios where the feature could benefit their work. While \p{1} did not find it valuable for their future writing, others identified uses such as determining paper structure (\p{2, 3, 4, 8, 9, 10, 11, 12}), outlining early drafts (\p{3, 4}), writing about unfamiliar subject areas or methods (\p{5, 6}),  writing for a new venue (\p{8}), having an \userquote{``external''} perspective (\p{11}), learning about research (\p{16}), preparing to break the norm in the research (\p{16}), prototyping an outline rapidly (\p{13}), and navigating when they have \userquote{``no idea where to start''} (\p{15}).}

\subsubsection{User Insights on Variation, Norms, and Breaking Norms}
\csnote{Participants noticed variation across section titles, noting that there were different ways to describe conceptually similar sections. Differences ranged from minor, such as singular versus plural titles (e.g., ``Related Work'' versus ``Related Works'') to more significant, like adding new details to a more generic title or the existence of a more distinct title indicating a more unusual paper component.
Evaluating the distribution of titles helped some participants see an unexpected range of possible titles and exposed them to new options. For example, \p{2} noted, \userquote{``The section titles are more diverse than I thought. There are things like, you know, `expert interviews' which I've never done... There were a few others that I found which are pretty new to me... It definitely broadens my sort of---my horizon.''}}

\csnote{Some participants were particularly interested in seeing titles that deviated from the norm. \p{3} found it helpful to visualize less common titles, which appeared only once or twice, compared to more popular ones with over a hundred instances. Some participants chose to use or adapt titles from the shown distribution, while others chose to intentionally deviate from the perceived norm of the distribution. For example, \p{7} decided to continue with their own drafting approach, as their paper was \userquote{``more on novel contribution rather than implementation,''} so they believed that frequently used section titles about implementation would not be useful when describing their work. 
Critically, we observed that participants exercised their agency despite having access to the distribution: If they did not like how a section title was phrased, even if it was frequently used, they would opt for their own desired phrasing.}

%% file: Sections/results/writing_strategies_with_retrieved_sentences.tex
\subsubsection{\csnote{Reading and Processing Sentences}}
\csnote{We observed that participants read and processed what they had read in two different styles: some did selective reading by skimming across sentences, while others read sentences in their entirety.}  \csnote{Regardless of engagement style, participants identified new takeaways.
	For example, \p{2} discovered new ways of structuring sentences through analyzing examples. 
	Some of the (sub)communities represented in the corpus were unfamiliar to \p{4} and \p{6}, and they learned about the habits reflected in that text. \p{14} likes control over the content of her writing, and valued learning about sentence structures that could guide her own composition choices.}

\cref{fig:user_phrases} shows two examples of system logs (not user interface screenshots) when participants retrieved sentences given a point in their draft. Note that the relevance of these retrieved results varied depending on each user's specific writing goals and context. %
\csnote{For example, several participants valued perceived syntax and organizational structures over content (\p{1, 2, 12, 10}).  \p{11} and \p{5} looked for concise sentences while \p{14} sought out more detailed examples.  \p{10} noted similarities in the retrieved sentences and reflected on drafting certain sections of their own.}

\subsubsection{Aligning, Understanding, and Addressing Terminology}
\csnote{Writing for different venues requires aligning terminology with community expectations to ensure consistency and clarity. Misunderstandings in terminology can make a paper difficult to follow or cause confusion, as \p{1} noted about differing definitions of terms like ``LLMs'' in different venues. \p{6} emphasized the importance of using correct terminology, especially in the analysis section, where there are expected norms to follow to avoid confusion. Participants found that they sometimes overlooked writing about important terms, as \p{11} realized when seeing the word ``rubric'' in the retrieved results, prompting him to address it in his manuscript. However, \p{6} cautioned that related terms might appear in different contexts, which requires careful consideration by the reader to avoid misinterpretation.}

\subsubsection{Finding Related Work}
\csnote{
Many participants used the interface as a tool to find related work to cite in their manuscript (\p{8,9,10,11,15,16}). Sometimes they intentionally searched for sources, while other times they came across works that highlighted a necessary concept or a limitation that they had not considered, prompting them to bookmark these sentences for later reference. \p{10} noted that they would use the tool more in the future to retrieve similar papers to cite and \p{11} would use it to find papers that included arguments related to their own work. \p{15} noted that they used the tool primarily for finding related work, not for identifying norms of a venue. In summary, the prototype demonstrated dual utility, serving not only as a writing aid for experienced and novice writers, but also as a research tool. Participants could develop their outline and manuscripts to align with, or break, expected norms and also discover related work for citation, all within one centralized platform.}

\subsubsection{Closing Content Gaps}
\csnote{Writers also used the tool to address content gaps in their drafts. 
\p{11} mentioned reading a sentence that helped her connect two key aspects of her work, confirming an idea she had previously considered but had not fully decided on. Another sentence prompted her to highlight a strength of her system in her manuscript. The sentences reminded participants of details they had forgotten to include, such as participant compensation (\p{13}). \p{15} used the sentences to identify gaps in their writing, stating, \userquote{``It's great to read what others wrote because we perform a similar process, but we did not mention that [...] in our own writing. So it would be great if we can add those.''}}

\subsubsection{Confidence and Perception}
\csnote{Participants felt more confident in their writing when they discovered that other authors had written about similar ideas or expressed them in a similar style. They also gained confidence in their related works section when they found papers in the retrieved results that aligned with their own literature review (\p{8, 15}). Two participants, after not finding  many related results for their initial text queries, perceived novelty in their writing topics and concluded that the community might not have discussed their topic yet. For example, \p{8} noted that his terminology appeared in other contexts but not his specific focus. \p{9} saw this as an opportunity to contribute novel insights. Similarly, \p{4} was encouraged by the lack of similar work, noting \userquote{``I'm also pretty happy that apparently this is a work that might not be yet that present [in] those two communities, or in the HCI community in general. So this is not something that I find bad, but actually something that encourages me, that I'm on the right track.''} Note that participants were informed at the beginning of the study that the corpus included papers from two venues over three years, so some relevant work may not have been included.}

%% file: Sections/results/results_tooltip.tex
Tooltips provided useful additional context (\p{3, 9, 13, 12}), enabling participants to contextualize retrieved sentences at the paper, section, and paragraph levels. The main critique was the additional text, which required extra effort to read.

At the paper level, the paper title offered clues about the overall topic, helping participants determine if a sentence was discussed in a similar context. For example, \p{9} commented that he \userquote{``ended up using that tooltip quite a lot. [...] because it was quite nice to directly make a connection from the sentence to where it came from.''} 

\csnote{When looking at the section level, participants found value in reviewing the section title path. \p{3} observed how other authors structured the formative study, introducing \userquote{``the design''} and \userquote{``the steps in the study''} before detailing the procedure. This led her to consider splitting her own formative study into subsections and adopting a less abrupt transition to better inform the system design.} 

At the paragraph level, participants reported that the sentences could sometimes be ambiguous when interpreted in isolation, so they read the surrounding sentences in the tooltip for clearer context. \p{12} used the tooltip \userquote{``to see what they're talking about before and after.''}

While the tooltips were helpful for many, three participants used it less frequently  (\p{2, 7, 14}). \p{7} commented, \userquote{``It just requires a lot of time to read it all [...] so I ended up not using it as much.''} \p{2} found that \userquote{``the two columns``} on the right hand side (i.e., \textit{What others wrote}, and \textit{What they wrote next}) provided enough context without needing the tooltip. Meanwhile, \p{14}, writing for a different venue, felt that HCI-focused contextualization was not relevant to her work.

%% file: Sections/results/writing_stages_utility.tex
Overall, participants found the design features useful both early in the writing process and during iterative revisions at later stages. 
Section title clusters were helpful for creating an initial outline. Three participants preferred to use it in early drafting because inspecting examples of other outlines helped them plan their writing for venues or contributions that they were unfamiliar with. 
Commenting on the integration of retrieved sentences, \p{1} mentioned that he would use them in the initial phase to establish terminology and sentence structure but not later, as the style, structure, and outline would already be set. Similarly, \p{10} preferred not to be influenced by other papers’ writing styles once his initial draft was more complete. \p{11} added that section title clusters and retrieved sentences are most helpful early on, when there are \userquote{``fewer restrictions''} on content and style. 

Once the style, terminology, and structure were set, e.g., when participants had written a complete paragraph, some participants shifted focus to refining their own writing rather than inspecting others’ work---but not all.
Two participants (\p{8, 11}) mentioned that they would use section title clusters in later stages to revise their outline, either to \userquote{``check the framing of the paper''} or adjust subsection titles. \p{3} suggested using the feature in different iterations of the paper to revise sections already written, explaining that she would \userquote{``highlight that section and compare it''}, via sentence retrieval, to similar work in the venue. Likewise, \p{4} noted that she would revisit the section titles, as the outline evolves during the writing process and may no longer align with the original version.

Some participants developed strategies to manage the complexity of all the features available to them. Often, they primarily focused on the \emph{What others wrote} column, consulting the \emph{What they wrote next} column and contextual tooltips only when needed for deeper understanding. \p{2} observed: \userquote{``I didn't look at the `What they wrote next' column that much... I also didn't look at the tooltips a lot, because I don't know. I was too absorbed by the 1st column---what others wrote. And because I,  at a given position or location in the paper... I was just trying to focus on what to write then.''} This selective attention may have helped them balance the information these features provided with the inherent cognitive load of lots of text that cognitive supports like special text rendering might not completely mitigate.

\subsection{Efficacy of Anti-plagiarism Features}

\csnote{We implemented measures intended to promote learning and prevent plagiarism.}
\csnote{As expected, while participants drew inspiration from the retrieved sentences, none showed an intent to plagiarize. Instead, they emphasized the importance of (i) modifying any examples to avoid plagiarism and align with their own ideas and (ii) saving work of interest to reference in their own paper, suggesting a strong awareness of the need to draft an original manuscript and cite properly.}

%% file: Sections/results/quantitative_results/task_outlining.tex
\subsection{Task 1: Outline Writing}
\begin{figure*}
	\centering
	\includegraphics[width=0.6\linewidth]{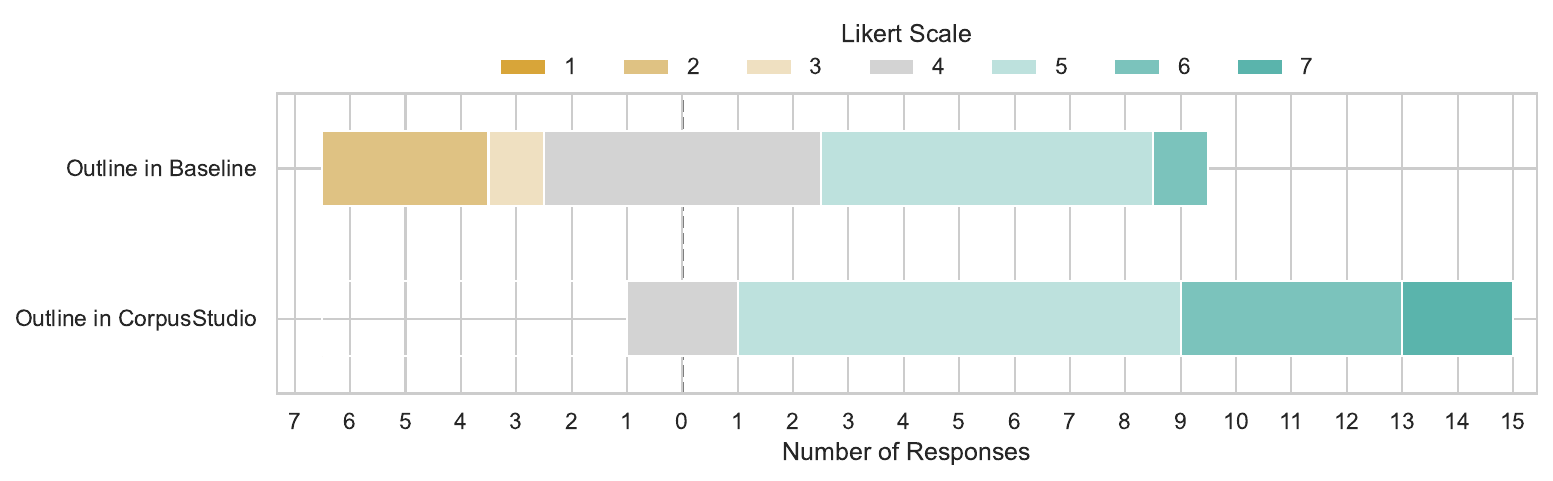}
	\caption{Usefulness of drafting an outline in the baseline condition compared to in \system}
	\Description{Answered on a
		seven point Likert scale from "Not useful at all" to "Very useful")}
	\label{fig:t1_doc_v_stop}
\end{figure*}

\subsubsection{Usefulness of Drafting Tools} 
Participants rated the usefulness of drafting their outline in a blank document (baseline) and drafting it in \system{} (\cref{fig:t1_doc_v_stop}). Results indicated a statistically significant preference for \system{} (p = 0.003). \csnote{As discussed in \Cref{sec:utility of section titles}, participants valued aspects such as creating a better structure for their outline, learning how to write for a new venue or topic, and confirming their existing titles.}

\begin{figure*}
	\centering
	\includegraphics[width=0.6\linewidth]{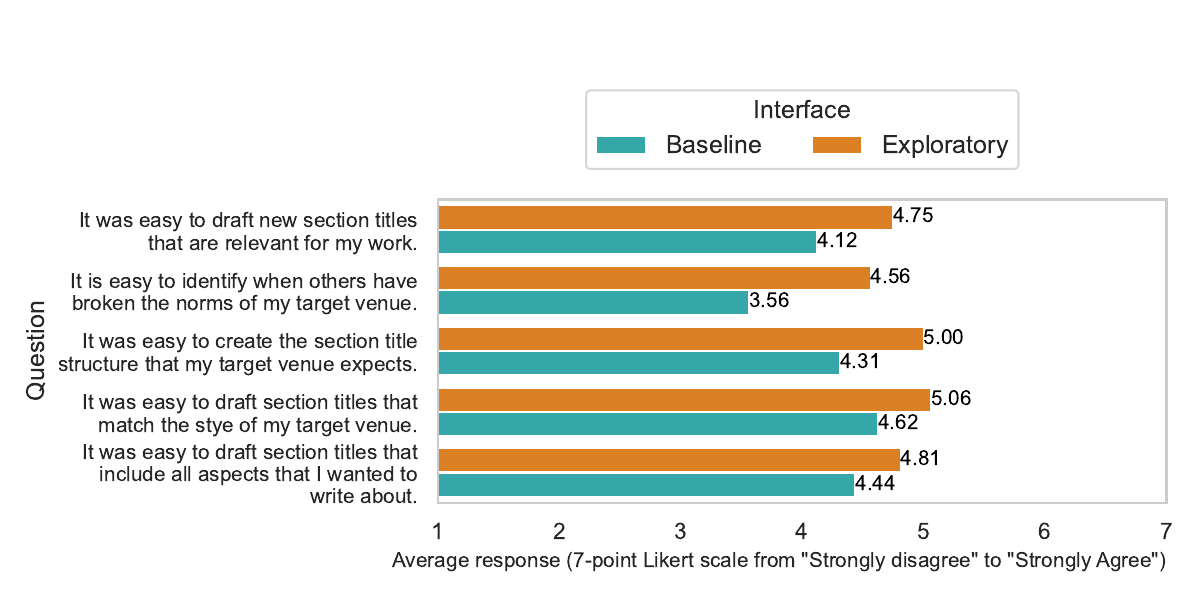}
	\caption{Ease of drafting an outline in the baseline condition compared to in \system{}.}
	\Description{Answered on a seven point Likert scale from "Strongly disagree" to "Strongly Agree"}
	\label{fig:likert_results_outline_drafting_easel}
\end{figure*}

\subsubsection{Ease of Drafting Section Title Outlines and Venue Alignment}
For both conditions, participants answered a series of questions focusing on how easily they drafted section titles that were relevant to their work, aligned with the style of the target venue, adhered to norms, included all desired aspects, \csnote{and identified when others had broken norms. The average responses for each condition are shown in \cref{fig:likert_results_outline_drafting_easel}. A Mann-Whitney U test revealed no statistically significant differences between the interface conditions for any of these five Likert items.}

\subsubsection{Differences Between Tasks}
Between the sub-tasks, i.e., drafting the first half of the document compared to the second half, there was a statistically significant difference (p=0.014) in participants' responses to the question, ``It was easy to create the section title structure that my target venue expects.'' This suggests potential learning effects \csnote{as participants found it easier to create the section title structure in the second half.} Participants could have adapted their writing strategies from drafting the first half to the second half. We found no significant differences in the other questions
within this category \csnote{(see \cref{app:t1-taskcompletion} for the full questions)}.

We also evaluated whether different interface conditions impacted the ease of drafting section titles and aligning them with the target venue's expectations for both the first and second halves of the outline. %
\csnote{Although no significant differences were found, our qualitative findings suggest that participants had greater confidence in their writing when using \system~ due to the availability of examples, which helped reduce uncertainty.}

\subsubsection{Participant Perceptions of Ease and Overwhelm}
Fourteen of the 16 participants (87.5\%) reported that writing the section titles in \system~ with \emph{Section Titles of Other Papers} made the task easier \csnote{(via alleviating the blank page problem and having examples at hand in a centralized writing environment)}, while the remaining two participants reported that writing in the baseline interface was easier. 

Nine of the 16 participants (56\%) concluded that writing section titles in \system~ felt more overwhelming \csnote{because they had to mentally process the distribution of section titles---yet most acknowledged that, despite the initial challenge, the process was useful for generating ideas.} The remaining seven participants (44\%) reported that completing the task in the baseline interface felt more overwhelming \csnote{because they had to begin from an empty page and they felt uncertainty about how to name and structure the section titles}.

\subsubsection{Identifying and Breaking Norms}
We evaluated whether there was any significant difference in participants' ability to identify norms of their target venue when reading past papers (prior to the study or during the study in the baseline condition) in comparison to reviewing the distribution over section titles. The results showed no significant difference. Similarly, for identifying when it is appropriate to break norms, no significant differences were found \csnote{(see \cref{app:t2-writingexpectations} for the full questions)}. 
\csnote{
	However, qualitative data highlighted that with \system, participants more frequently reflected on breaking norms intentionally, using the example distributions and their own writing to justify deviations. While these reflections did not result in significant quantitative differences, they suggest a deeper engagement with norms during the task.
}

%% file: Sections/results/quantitative_results/task_writing.tex
\subsection{Task 2: Writing a Section of a Manuscript}

\begin{figure*}
	\centering
	\includegraphics[width=0.7\linewidth]{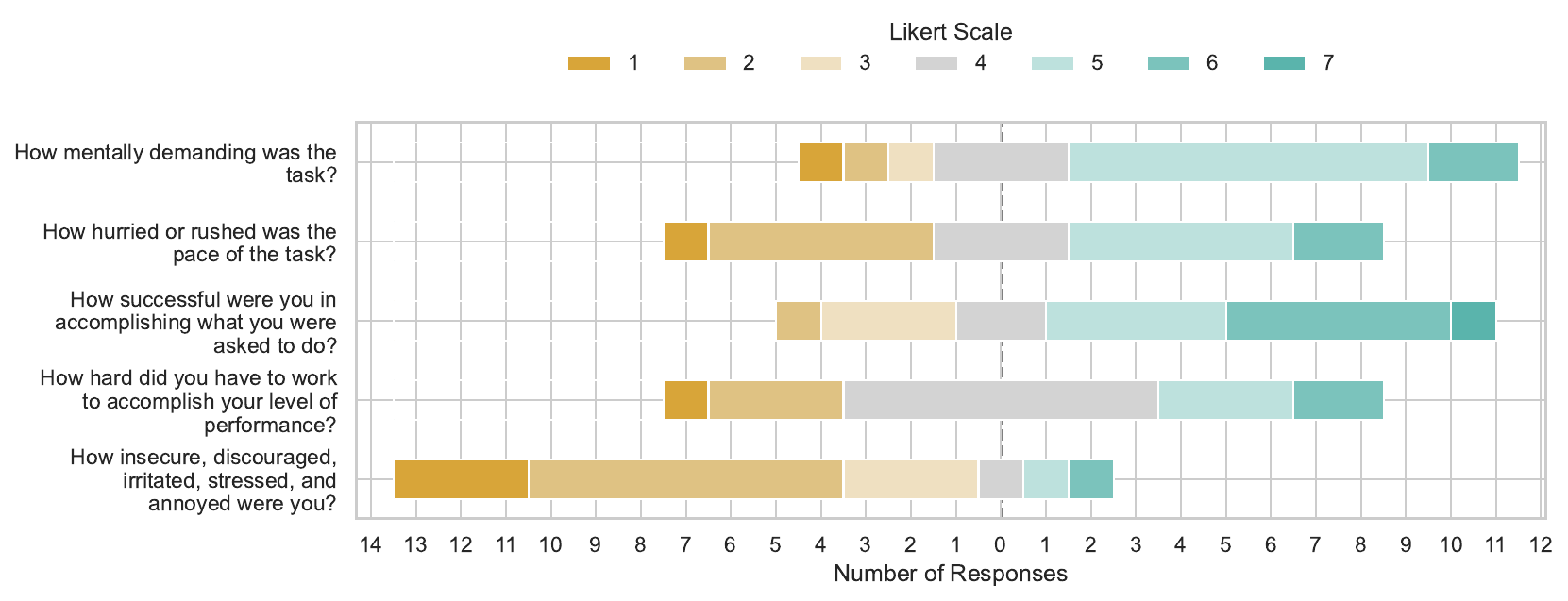}
	\caption{Responses to the NASA-TLX questionnaire during Task 2: writing a section of a manuscript}
	\Description{A Plot showing participants' responses to the NASA-TLX questionnaire}
	\label{fig:nasa_tlx}
\end{figure*}

\begin{figure*}
	\centering
	\includegraphics[width=0.7\linewidth]{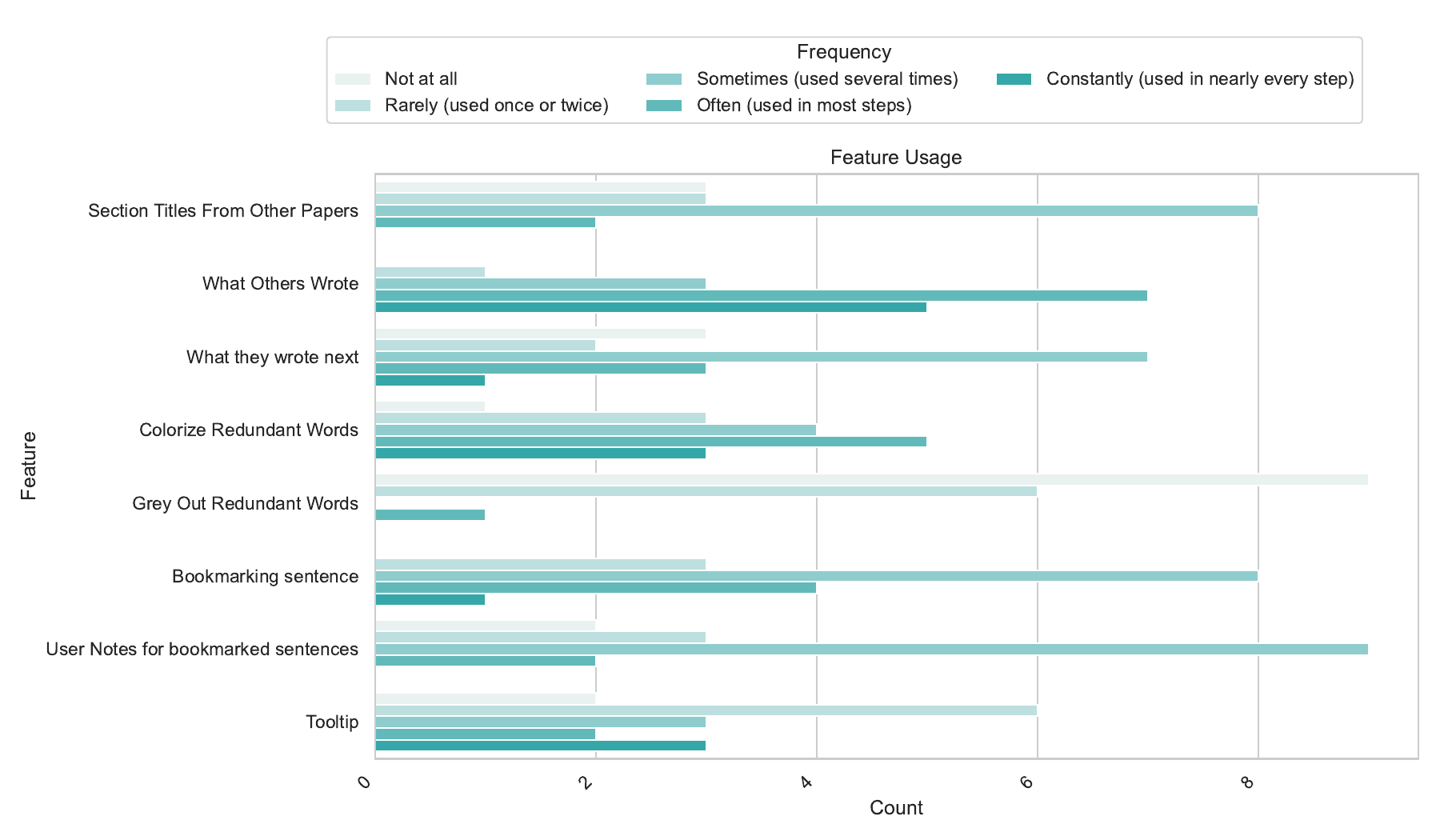}
	\caption{Participant-reported frequency of using each feature while completing the writing task}
	\Description{Plot showing how often participants used each design feature during the writing task.}
	\label{fig:feature_freq}
\end{figure*}

\begin{figure*}
	\centering
	\includegraphics[width=0.7\linewidth]{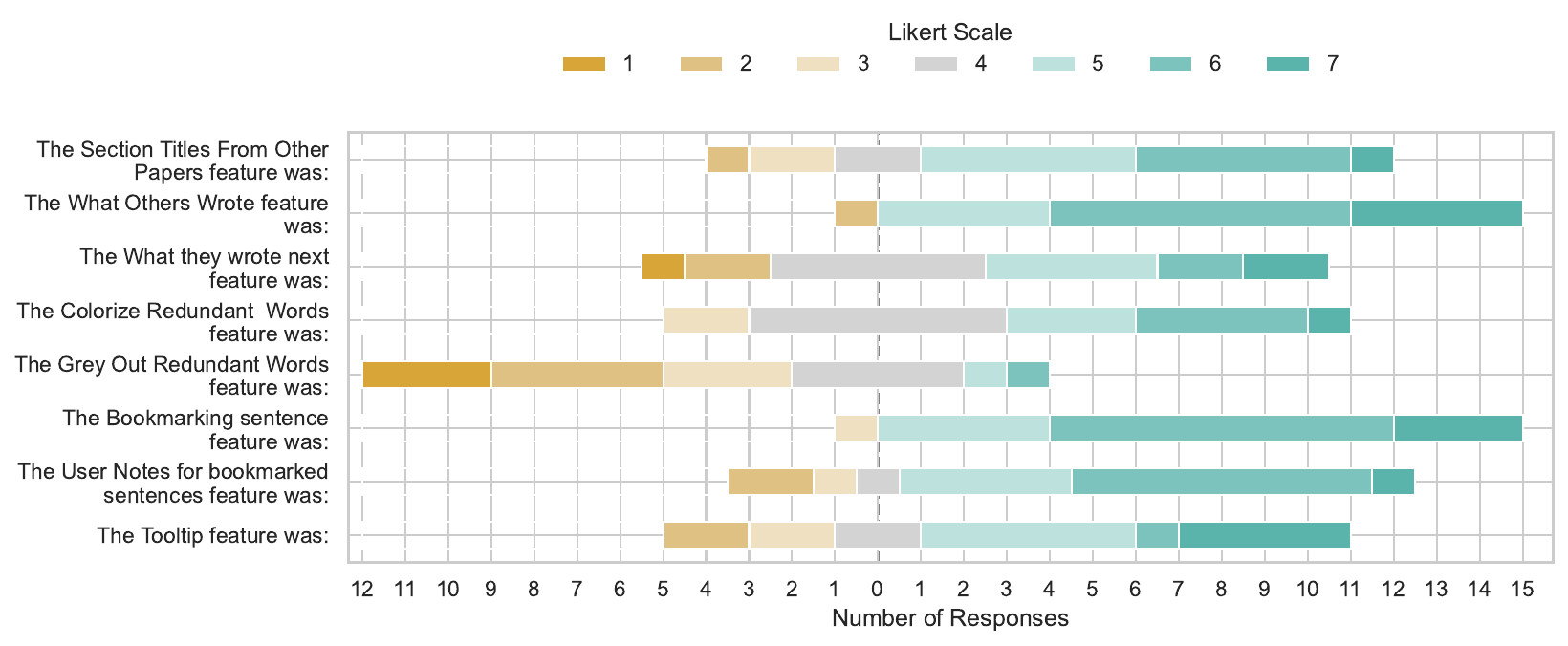}
	\caption{Usefulness of each feature during the drafting process}
	\Description{Plot listing the each feature and a Likert-scaled diverging bar chart of participants' responses.}
	\label{fig:feature_useful}
\end{figure*}

Based on the NASA-TLX questionnaire (\cref{fig:nasa_tlx}), %
participants found the writing task mentally demanding, which balanced with their perception of how hard they had to work. They still felt successful in completing the task. %

Participants reported on how often they used each feature (\cref{fig:feature_freq}), with the counts reflecting total reported usage across all participants while completing the writing task. The retrieved sentences in the \emph{What others wrote} column was most frequently used, %
followed by the \emph{colorizing redundant words} mode of rendering sentences, which participants reported using for reasons including finding relevant aspects, remaining focused, reading faster, and skimming text (\cref{sec:visual_support}). They also frequently used the \emph{tooltips}, noting that they were useful for contextualizing retrieved sentences and identifying their source (\cref{sec:tooltip}).  \emph{Greying out repeated words} was least frequently used. Some participants did not find it useful to de-emphasize words, while others did not notice the feature at all (\cref{sec:visual_support}). %
There was an overall positive sentiment on the usefulness of all other features (\cref{fig:feature_useful}).

\begin{figure*}
	\centering
	\includegraphics[width=0.7\linewidth]{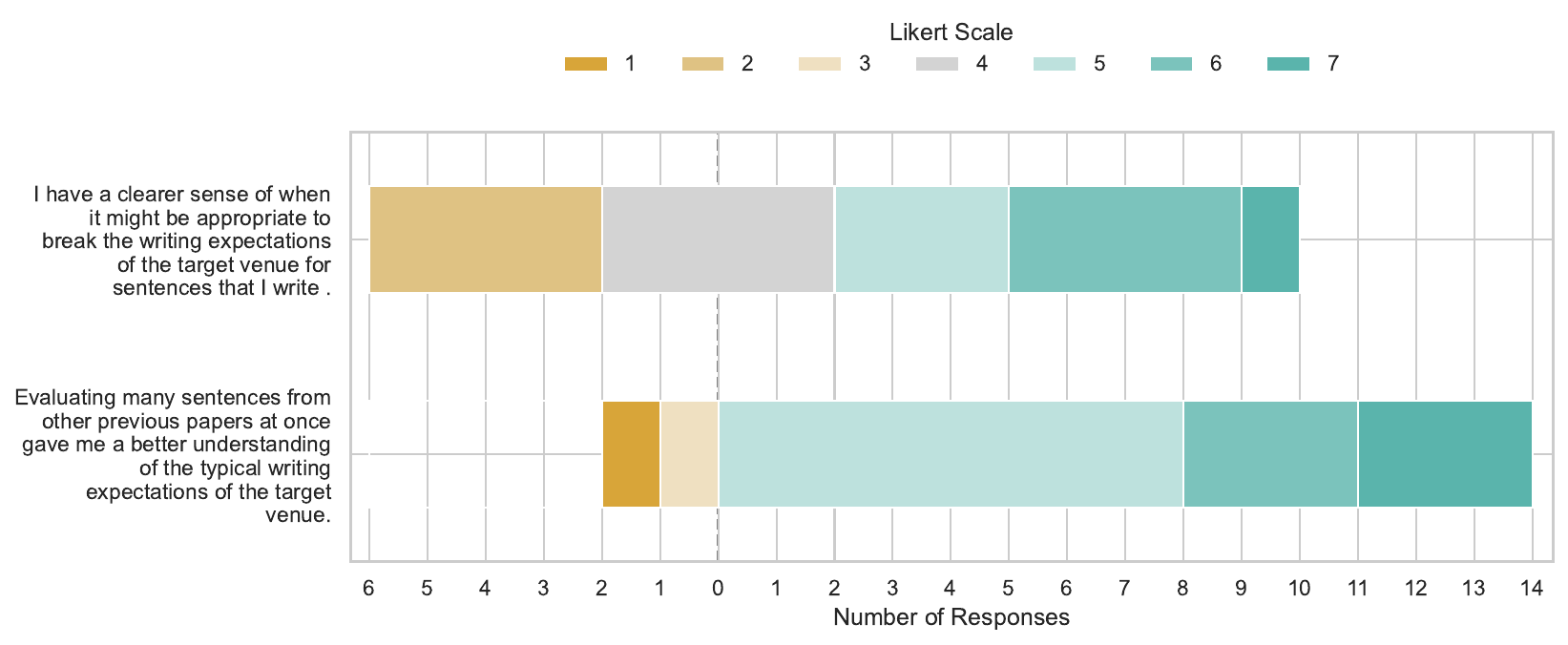}
	\caption{Perceptions on understanding and breaking typical writing expectations of the target venue after using \system}
	\Description{A plot with two questions highlighting whether participants writing expectations and understanding of the community's writing conventions changed.}
	\label{fig:t2_expectations}
\end{figure*}

Eight of 16 participants (50\%) answered that they had a clearer sense of when it was appropriate to break the writing expectations of the target venue after interacting with \system~ while four did not think that their understanding had changed and four gave a neutral response (\cref{fig:t2_expectations}).
From the study, we observed that at least three people commented on deviations between their prior assumptions about the structure or content and the retrieved content (\cref{sec:utility of section titles}).

All but two participants found that evaluating many sentences from previous papers gave them a better understanding of the typical writing expectations of the target venue (\cref{fig:t2_expectations}). This aligns with the qualitative results where participants commented positively about the usefulness of inspecting retrieved sentences, which gave them more confidence in their writing.

%% file: Sections/discussion.tex
\section{Discussion}

Our analysis revealed several ways that writers used \system{} to support their academic writing process. Here, we reflect on our design and results on conceptual and empirical levels.

\subsection{Relevance of Retrieved Sentences}
\changenote{
	We found that relevance of retrieved sentences was tightly coupled with users' writing intentions, as they found different uses and rationales for integrating the retrieved results into their writing process.
}

\subsubsection{Users' Diverse Applications of Retrieved Sentences}
\changenote{
	While \system{} uses text embeddings to capture semantic and syntactic relationships \cite{muennighoff_2023_mteb}, our study revealed that relevance was also determined by participants' writing goals rather than semantic similarity alone. For instance, semantically similar sentences were sometimes less useful than structurally similar ones when writers sought to understand the organization of the paper.
}

\changenote{
	Participants sought inspiration at multiple levels of writing: from paper organization through section titles to section-level transitions and standard academic phrases. The writers shaped these results through queries using their draft text itself, similar to patterns observed in interactions with AI in writing \cite{Dang2023ChoiceOC}. This challenges fundamental information retrieval assumptions about query-document similarity as an (only) indicator of relevance. While it provided a useful starting point, our findings suggest that relevance of retrieved text in writing support systems is not fixed, but dynamically constructed based on writers' immediate goals. 
}

\subsubsection{Interpreting Retrieved Results: User Behaviors and Dataset Limitations}
\changenote{
	Our study revealed how retrieved examples influenced participants' writing in unexpected ways. %
	When participants were looking for something specific and did not find it in the retrieved results, this led to two notable behaviors: First, participants often rephrased their text to get different results, blurring the line between writing and querying. Second, some interpreted a lack of results as evidence of their work's novelty, a problematic conclusion given \system's limited scope (two HCI venues: IUI and UIST from 2021-2023).
}
\changenote{
	These findings suggest that writing support systems should actively communicate their scope and limitations, for instance by suggesting broader venue searches when topic-specific queries yield no results, helping writers make informed decisions about how to interpret search results.
}

\subsection{Designing a System with Many Retrieved Examples}

\changenote{
	While many writing support systems prioritize simplicity through aggregated results, we deliberately presented many examples (N=25) to increase the variety of examples users saw. To help users process the larger scale of information, we provided sentence rendering modes that modulate visual variables of the text in a way intended to support users in recognizing cross-sentence relationships. While it provided support, it did not eliminate all the additional cognitive demands of the extra information.
}

\changenote{
	The study results suggest that showing multiple examples at once is valuable despite its cognitive demands: seeing variations in how similar ideas were expressed helped participants distinguish between rigid conventions and flexible patterns in academic writing. This suggests a trade-off between cognitive load and learning potential, a consideration for future writing support systems that aim to support learning rather than just text production. 
	
	Given that readers have different visual preferences, we provided more than one method for cognitive support. We encourage designers who wish to show multiple parallel examples to also provide alternate ways to help users visually identify patterns.
}

\subsection{Designing Writing Support for and with Community Integration, not AI Integration}

Our concept provides relevant text snippets to support writing but it does not use any AI-generated text, in contrast to recent trends in writing assistant systems~\cite{Lee2024designSpace}. Instead of training AI on a corpus, we surface examples from the corpus directly. In our case, these examples can be assumed to meet the community's quality standards, because they are evaluated by its members beforehand, that is, the published papers in the corpus went through peer review.

We argue that this approach values and supports communities. For example, it might aid communities in discovering their own community-level patterns of expression and in ``onboarding'' new community members. Supporting this, our design includes functionality for taking one's own notes, which might later contribute to exchanges between community members about writing norms, i.e., beyond individual interactions with our system.

This approach may also make a community's past work more discoverable to its members. Our retrieval method is based on the local text in the draft, and not on other metrics, such as paper citation counts or author metrics. In this way, our concept might contribute to discovering less well known but relevant work that the writer would not have noticed otherwise. This could be examined in a dedicated future study, and the UI design could be developed further in this direction.

\subsection{Putting Metacognitive Strategies ``Ready to Hand''} 
\changenote{
	\citet{Tankelevitch2024metacognitive} recently motivated designing for metacognitive challenges in interaction. One emerging approach here is to do so by embedding specific thinking strategies directly into the interface.
}
For instance, \citet{Dang2022beyondTextGen} designed for the strategy of ``reverse outlining'' by providing a sidebar of paragraph-wise summaries, updated live. Similarly, we designed for learning from examples, with respective material provided in sidebars directly within the writing environment. Our design also  
supports the externalization of users' self-reflection (e.g., what does this mean for my text?) in User Notes.
As our results show, participants found this helpful and made use of it in diverse ways. 

\citet{Tankelevitch2024metacognitive} also suggests ``support[ing] users in increasing their self-awareness and adjusting their confidence.'' We observed that, by becoming aware of what authors in a community wrote, many participants' confidence levels increased with respect to writing for that same community.

\subsection{Limitations and Future Work}
Our results reveal how writers interact with and perceive examples retrieved from a community-specific document corpus. However, readers should be cautious when generalizing these findings beyond the scope of our study. All participants had a background in HCI, and the corpus was drawn from two HCI venues, focusing primarily on system contribution papers. Additionally, the writing task was based on an academic writing scenario. Future research should explore other writing contexts where communities play a role, such as fiction, poetry, or argumentative writing.
In our user study, participants were given 20-25 minutes to draft their text and engage with the various system features. While this offered valuable insights into their interactions and perceptions of the writing environment, it would be worthwhile to explore how prolonged use of these features might influence users’ writing habits and style.

%% file: Sections/conclusion.tex
\section{Conclusion}
\changenote{\system{} is a writing support system that integrates example retrieval---at both the document level and sentence level---directly into the drafting process. Writers can compare their text against a curated corpus from their target academic community, with the system presenting many real examples in parallel. This tight coupling between writing and retrieval enables authors to identify established writing patterns, understand community conventions, and validate their writing choices through direct comparison with published work.}
\changenote{Unlike systems that generate or remix text, \system{} deliberately preserves and attributes original manuscripts, using search based on text embeddings that incorporate spatial information and context to surface relevant examples in an accessible format. Thereby we support academic writing while maintaining transparency and acknowledging existing scholarship. Our work aims to advance an interaction model where writing support tools help authors overcome writing barriers while respecting and learning from the contributions of others in their field.}

%% file: Sections/Appendix.tex
\section{Surveys} \label{app: surveys}
The surveys we used in the study.
\subsection{Pre-study Survey}
\label{app:pre-study}

\subsubsection{Demographics}
\begin{enumerate}
    \item What is your participant ID?	\newline
    (Given to participant by interviewee)
    \item What is your age? \newline
        \begin{itemize*}%
            \item[$\circ$] 18-24
            \item [$\circ$]25-34
            \item [$\circ$]35-44
            \item [$\circ$]45-54
            \item [$\circ$]54+
            \item[$\circ$] Prefer not to disclose
        \end{itemize*}
    \item What is your gender?	\newline
        \begin{itemize*}%
            \item [$\circ$]Woman
            \item[$\circ$] Man
            \item [$\circ$]Non-binary
            \item[$\circ$] Prefer not to disclose
            \item[$\circ$] Other:
        \end{itemize*}
    \item What is your profession? \newline
        \begin{itemize*}%
            \item [$\circ$]Undergraduate student 
            \item[$\circ$] Masters student 
            \item [$\circ$]PhD student/candidate
            \item [$\circ$]Faculty
            \item [$\circ$]Other: 
        \end{itemize*}
\end{enumerate}

\subsubsection{Need for Cognition (NCS-6) scale} 
The following questions are on a 7 point likert scale from ``not at all like me'' to ``very much like me.''
\begin{enumerate}
    \item I would prefer complex to simple problems. 
    \item I like to have the responsibility of handling a situation that requires a lot of thinking.
    \item Thinking is not my idea of fun.
    \item I would rather do something that requires little thought than something that is sure to challenge my thinking abilities.
    \item I really enjoy a task that involves coming up with new solutions to problems. 
    \item I would prefer a task that is intellectual, difficult, and important to one that is somewhat important but does not require much thought. 
\end{enumerate} 

\subsubsection{Writing Experience and Approach}
\begin{enumerate}
    \item What is your field of study? \newline
    (Open-ended)
    \item What is the topic for the paper you are working on today? \newline
    (Open-ended)	
    \item What is your target venue for this writing?	\newline
    (Open-ended)
    \item How many years of experience do you have with academic writing? Use a numeric answer. \newline
    (Open-ended)	
    \item How many years of experience do you have with writing for your selected target venue? Use a numeric answer.\newline
    (Open-ended)	
    \item How familiar are you with the process of writing for your selected target venue?	
    \item How many academic papers have you authored or co-authored? Use a numeric answer.	\newline
    (Open-ended)
    \item How confident are you in identifying the writing style and content expected by a particular academic venue?	\newline
    (Answered on a 7-point likert scale from ``Not confident at all''  to ``Extremely confident'')
    \item Have you used any writing support tools (e.g., Grammarly, Ref-n-Write, SciSpace) in the past?	\newline
        \begin{itemize*}%
            \item[$\circ$] Yes
            \item [$\circ$]No
        \end{itemize*}
    \item What writing support tools have you used in the past? \newline
    (Open-ended)
    \item How often do you use writing support tools? \newline	             \begin{itemize*}%
            \item[$\circ$] Never
            \item [$\circ$]I've used them a few times, but not regularly
            \item [$\circ$]A few time a month
            \item [$\circ$]A few times a week
            \item[$\circ$] Once a day or more
        \end{itemize*}
    \item What strategies do you typically use when drafting a section of an academic paper?	\newline
    (Open-ended)
    \item What strategies do you typically use when drafting a section of a paper for your selected target venue? \newline
    (Open-ended)	
    \item How often do you refer to published papers while writing? \newline
        \begin{itemize*}%
            \item [$\circ$]Never
            \item [$\circ$]Rarely
            \item [$\circ$]Sometimes
            \item [$\circ$]Often
            \item [$\circ$]Always
        \end{itemize*}
    \item What challenges do you usually face when writing for an academic venue?	\newline  
    (Open-ended)
\end{enumerate}

\subsection{Task 1 Outlining Survey - Page without section titles}
\label{app:t1-baseline}
\subsubsection{Participant and Task Identification}
\begin{enumerate}
    \item What is your participant ID?	\newline
    (Given to participant by interviewee)
    \item What outlining task did you just complete? \newline
        \begin{itemize*}%
            \item [$\circ$]First half of manuscript
            \item[$\circ$] Second half of manuscript
        \end{itemize*}
\end{enumerate}
\subsubsection{Interface Comparison}
Participants answered these questions after outlining the second half of the manuscript.
\begin{enumerate}
    \item Which interface made the task easier?	\newline
        \begin{itemize*}%
            \item [$\circ$]Page with section titles
            \item [$\circ$]Page without section titles
        \end{itemize*}
    \item Which interface felt more overwhelming? \newline
        \begin{itemize*}%
            \item [$\circ$]Page with section titles
            \item [$\circ$]Page without section titles
        \end{itemize*}	
\end{enumerate}
\subsubsection{Task Completion}
The following questions are on a 7 point likert scale from ``Strongly disagree'' (1) to ``Strongly agree'' (7):
\begin{enumerate}
    \item It was easy to draft section titles that include all aspects that I wanted to write about.	
    \item It was easy to draft section titles that match the stye of my target venue.	
    \item It was easy to create the section title structure that my target venue expects.	 
    \item It is easy to identify when others have broken the norms of my target venue.	
    \item It was easy to draft new section titles that relevant for my work.	
\end{enumerate}
\subsubsection{Writing Expectations}
\begin{enumerate}
    \item What is your current strategy for comparing paper outline from related papers \newline
    (Open-ended)
    \item When I looked at previous papers from my target venue in the past, evaluating many section titles from those papers gave me a better understanding of the typical writing expectations of the target venue. \newline 
    7 point likert scale from ``Strongly disagree'' (1) to ``Strongly agree'' (7)
    \item Based on when I looked at previous papers from my target venue in the past, I have a clearer sense of when it might be appropriate to break the writing expectations of that target venue for section titles that I write. \newline 7 point likert scale from ``Strongly disagree'' (1) to ``Strongly agree'' (7)
\end{enumerate}
\subsubsection{Feature Usefulness}
The following question is on a 7 point likert scale from ``Not useful at all'' (1) to ``Very useful'' (7):
\begin{enumerate}
    \item Writing in the document was:
\end{enumerate}

\subsection{Task 1- Page with section titles}
\label{app:t1-exploratory}
\subsubsection{Participant and Task Identification}
\begin{enumerate}
    \item What is your participant ID?	\newline
    (Given to participant by interviewee)
    \item What outlining task did you just complete? \newline
        \begin{itemize*}%
            \item[$\circ$] First half of manuscript
            \item [$\circ$]Second half of manuscript
        \end{itemize*}
\end{enumerate}
\subsubsection{Task Completion}
\label{app:t1-taskcompletion}
The following questions are on a 7 point likert scale from ``Strongly disagree'' (1) to ``Strongly agree'' (7):
\begin{enumerate}
    \item It was easy to draft section titles that include all aspects that I wanted to write about.	
    \item It was easy to draft section titles that match the style of my target venue.	
    \item It was easy to create the section title structure that my target venue expects.	 
    \item It is easy to identify when others have broken the norms of my target venue.	
    \item It was easy to draft new section titles that relevant for my work.	
\end{enumerate}
\subsubsection{Writing expectations}
The following questions are on a 7 point likert scale from ``Strongly disagree'' (1) to ``Strongly agree'' (7):
\begin{enumerate}
     \item Evaluating many section titles from other papers at once gave me a better understanding of the typical writing expectations of the target venue.	
    \item After evaluating many section titles from other papers at once, I have a clearer sense of when it might be appropriate to break the writing expectations of the target venue for section titles that I write.	
\end{enumerate}
\subsubsection{Feature Usefulness}
The following question is on a 7 point likert scale from ``Not useful at all'' (1) to ``Very useful'' (7):
\begin{enumerate}
    \item The Section Titles From Other Papers feature was:
\end{enumerate}
 
\subsection{Task 2 - Post-writing Survey}
\label{app:t2-survey}
\subsubsection{Participant identification}
\begin{enumerate}
    \item What is your participant ID?	
\end{enumerate}
\subsubsection{NASA-TLX}
The following questions from the \emph{NASA-TLX)} scale are on a 7 point likert scale from ``Very low'' (1) to ``Very high'':
\begin{enumerate}
    \item How mentally demanding was the task?	
    \item How hurried or rushed was the pace of the task?	
    \item How successful were you in accomplishing what you were asked to do?	
    \item How hard did you have to work to accomplish your level of performance?	
    \item How insecure, discouraged, irritated, stressed, and annoyed were you?	
\end{enumerate}
\subsubsection{Writing expectations}
\label{app:t2-writingexpectations}
The following question is on a 7 point likert scale from ``Strongly disagree'' (1) to ``Strongly agree'' (7):
\begin{enumerate}
    \item Evaluating many sentences from other previous papers at once gave me a better understanding of the typical writing expectations of the target venue.	
    \item I have a clearer sense of when it might be appropriate to break the writing expectations of the target venue for sentences that I write.	
\end{enumerate}
\subsubsection{Feature Evaluation}
For each feature, participants ranked the usefulness on a 7-point likert scale from ``Not useful at all'' (1) to ``Very useful'' (7); we included a screenshot of the respective feature for participants to reference when answering the question. They then reported their frequency of using the feature for completing the task with the following options: Not at all, Rarely (used once or twice), Sometimes (used several times), Often (used in most steps), and Constantly (used in nearly every step). Lastly, they answered an open-ended question on scenarios for which the feature would be most valuable in their future writing tasks. The questions for each feature are listed below.
\begin{enumerate}
    \item The Section Titles From Other Papers feature was:	
    \item How frequently did you use the Section Titles From Other Papers feature while completing the task?	
    \item In what scenarios do you see the Section Titles From Other Papers feature being most valuable in your future writing tasks?	
    
    \item The What Others Wrote feature was: 	
    \item How frequently did you use the What Others Wrote feature while completing the task?	
    \item In what scenarios do you see the What Others Wrote feature being most valuable in your future writing tasks?	
   
    \item The What they wrote next feature was: 	
    \item How frequently did you use the What they wrote next feature while completing the task?	
    \item In what scenarios do you see the What they wrote next feature being most valuable in your future writing tasks?	
    
    \item The Colorize Redundant Words feature was: 	
    \item How frequently did you use the  Colorize Redundant Words feature while completing the task?	
    \item In what scenarios do you see the  Colorize Redundant  Words feature being most valuable in your future writing tasks?	
    
    \item The Grey Out Redundant Words feature was:	
    \item How frequently did you use the  Grey Out Redundant Words feature while completing the task?	
    \item In what scenarios do you see the  Grey Out Redundant Words feature being most valuable in your future writing tasks?	
    \item The Bookmarking sentence feature was:	How frequently did you use the Bookmarking sentence feature while completing the task?	
    \item In what scenarios do you see Bookmarking sentence feature being most valuable in your future writing tasks?	
    
    \item The User Notes for bookmarked sentences feature was:	
    \item How frequently did you use the User Notes for bookmarked sentences feature while completing the task?	
    \item In what scenarios do you see User Notes for bookmarked sentences feature being most valuable in your future writing tasks?	
    \item The Tooltip feature was:	How frequently did you use the Tooltip feature while completing the task?	
    \item In what scenarios do you see Tooltip feature being most valuable in your future writing tasks?				
\end{enumerate}

\subsection{Semi-structured Interview}
After participants completed both tasks and all surveys, we asked participants the following questions:
\label{app:interview}
\begin{enumerate}
    \item How was your experience using these features?
    \item What was your approach when writing with the different features?
    \item Now that you have a better understanding of the different features, when and
    where would you see the features have the greatest impact on your writing
    process?
    \item Did you face any challenges while writing this section?
    \item Imagine you didn't have this tool, and think back to your experiences today. What
    approach would you have taken if you wanted to accomplish the same thing, but didn't
    have this interface?
    \item For each named feature (Section Title Explorer, Colorize similar words, Grey Out Redundant Words, Tooltip, Bookmarking sentences, User Notes)
            \item What did you like about this feature?
            \item What did you not like about this feature?
            \item What do you wish it could do?
            \item Would you use this feature in the future? If so, how and for what purpose?
            If not, why not?
    \item Has your understanding of the writing norms of your target community changed?
    And if so how?
    \item Outside of the discussed features, are there any recommendations you'd make for improving
    the tool?
    \item Is there anything that I did not ask about that you want to share?
\end{enumerate}

\subsection{Additional Figures}

 \begin{figure}
    \centering
    \includegraphics[width=8cm]{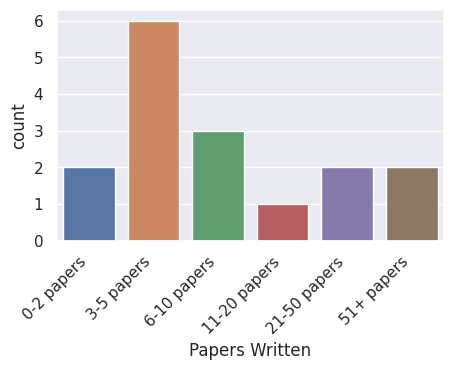}
    \caption{``How many academic papers have you authored or co-authored?''}
    \Description{Plot showing participants' reported number of papers written}
    \label{fig:papers_written}
 \end{figure}